\documentclass[onecolumn]{aastex6}
\usepackage{threeparttable}

\newcommand{\kms}{km~s$^{-1}$}

\newcommand{\cii}{C\,{\sc ii}}

\newcommand{\civ}{C\,{\sc iv}}

\newcommand{\lala}{$\lambda$$\lambda$}

\newcommand{\lya}{Ly$\alpha$}

\newcommand{\oi}{O\,{\sc i}}

\newcommand{\siv}{Si\,{\sc iv}} 

\newcommand{\SIii}{Si\,{\sc ii}}

\newcommand{\asec}{$^{\prime\prime}$}

\newcommand{\BI}{BI$^{*}$}

\begin{document}

\title{Emergence and Variability of Broad Absorption Line Quasar Outflows}

\author{J. A. Rogerson}
\affil{Canada Aviation and Space Museum, Ottawa, ON K1K 4Y5, Canada\\
Department of Physics and Astronomy, York University, Toronto, ON M3J 1P3, Canada
}

\author{P. B. Hall, N. S. Ahmed}
\affil{Department of Physics and Astronomy, York University, Toronto, ON M3J 1P3, Canada}

\author{P. Rodr\'{\i}guez Hidalgo}
\affil{Department of Physics \& Astronomy, Humboldt State University, Arcata, CA 95521, USA\\
Department of Physics and Astronomy, York University, Toronto, ON M3J 1P3, Canada}
\author{W. N. Brandt}
\affil{Department of Astronomy \& Astrophysics, 525 Davey Lab, The Pennsylvania
State University, University Park, PA 16802, USA \\
Institute for Gravitation and the Cosmos, The Pennsylvania State University,
University Park, PA 16802, USA \\
Department of Physics, 104 Davey Lab, The Pennsylvania State University,
University Park, PA 16802, USA}
\and
\author{N. Filiz Ak}
\affil{Faculty of Sciences, Department of Astronomy and Space Sciences, Erciyes University, 38039 Kayseri, Turkey \\
Astronomy and Space Sciences Observatory and Research Center, Erciyes University, 38039 Kayseri, Turkey}

\begin{abstract}
We isolate a set of quasars that exhibit emergent \civ\ broad absorption lines (BALs) in their spectra
by comparing spectra in the SDSS Data Release 7 and the SDSS/BOSS
Data Releases 9 and 10.
After visually defining a set of emergent BALs, follow-up
observations were obtained with the Gemini Observatory for 105 quasars.
We find an emergence rate consistent with
the previously reported disappearance rate of BAL quasars
given the relative numbers of non-BAL and BAL quasars in the SDSS.
We find candidate newly emerged BALs are preferentially drawn from among BALs with smaller balnicity indices, shallower depths, larger velocities, and smaller widths.
Within two rest-frame years (average) after a BAL has emerged,
we find it equally likely to continue increasing in equivalent width in an observation six months later (average) as it is to start decreasing.
From the time separations between our observations, we conclude
the coherence time-scale of BALs is less than 100 rest-frame days.
We observe coordinated variability among pairs of troughs in the same quasar, likely due to clouds at different velocities responding to the same changes in ionizing flux; and the coordination is stronger if the velocity separation between the two troughs is smaller.
We speculate the
latter effect may be due to clouds having on average lower densities at higher velocities due to mass conservation in an accelerating flow, causing
the absorbing gas in those clouds
to respond on different timescales to the same ionizing flux variations.
\end{abstract}


\section{Introduction} \label{sec:intro}

Several types of absorptions are found in the spectra of quasars, including
narrow and broad intrinsic absorption. Narrow absorption lines
(NALs) are narrow enough that the doublet lines from the species (e.g., \civ, \siv, etc.) are resolved. Broad absorption lines (BALs) occur
in the same species, but over a large enough velocity regime that the doublets
are blended together.

The first broad absorption troughs were identified by \citet{lyn67} who noticed
wide and strong absorption in \civ\ and \siv\ blueshifted from their respective
emission features.  That shift to shorter wavelengths
means that the gas that is absorbing the light
must be moving away from the quasar and toward Earth.
%
BAL absorption troughs are most often seen to absorb only some fraction of the continuum light, meaning that the absorbing gas either covers only part of the continuum source, or that the gas completely covers the continuum source but is not optically thick, or both.

Conventionally,
broad absorption Line (BAL) quasars have been defined as quasars
that exhibit blueshifted absorption due to the \civ\ doublet at \lala\
1548.203, 1550.770 \AA\ that is at least
2,000~\kms\ wide and can extend from 3,000~\kms\
to 25,000~\kms, where 0~\kms\ is at the systemic redshift of
the quasar \citep{wea91}. Modifications to this
definition have been proposed -- e.g., \citet{sdss123} and \citet{trump06} --
to include absorption features that excluded by the original definition.
Regardless
of the exact definition, broad absorption is rooted in a physically distinct
origin compared to intervening absorption features in quasar spectra.

The standard disk-wind picture for BAL outflows
consists of a supermassive black hole, surrounded by a relatively thin
accretion disk with a UV-emitting region at small radii, and BAL features
arising from material lifted off the accretion disk and
accelerated at least in part by radiation line driving to high outflow
velocities that we observe as blueshifted absorption (e.g., \citealt{mcgv},
\citealt{elv00}, \citealt{pk04}, \citealt{2010ApJ...722..642O}). In this model we are
seeing primarily the wind's radial motion as it is accelerated away from the
central source.  It also means our line-of-sight affects whether we see
a BAL or not.
Thus, observing a quasar outflow provides insight into the structure of the
central engine. Outflows of this magnitude may also represent a mechanism
by which supermassive black holes provide feedback to their host galaxy (e.g.,
\citealt{2009ApJ...706..525M},
\citealt{2013MNRAS.436.3286A},
\citealt{2014ApJ...788..123L},
\citealt{2015MNRAS.450.1085C}).

The variability of BALs may provide even more insight into
the structure of the central engine \citep{pprhh}.
The first multi-object sample of BAL variability was
published by \citet{bar94}, who collected a set of 23 BAL quasars and observed
them at least twice.
Variability in the strength (i.e., the depth, width, or outflow velocity profile)
of BALs is now a well documented phenomenon both in individual quasars and in large samples; see Table~\ref{tab:VarLit} for a non-exhaustive list of multi-object \civ\ variability studies currently in the literature.
There have been recent studies documenting the disappearance of BAL
troughs (e.g., \citealt{2012ApJ...757..114F}) as well as
BAL emergence in quasars that were not
classified as having BALs previously (e.g., \citealt{2008MNRAS.391L..39H}, \citealt{2009ApJ...701..176L},
\citealt{2010ApJ...724L.203K}, \citealt{J0230},
\citealt{2017arXiv170503019M},
Rodr\'{i}guez Hidalgo et al., in preparation).
This behavior indicates that our ability to observe broad absorption lines in quasars can depend on local factors as well as on our viewing angle.

\begin{table}
  \begin{threeparttable}
    \caption{Summary of BAL Quasar \civ\ variability studies in the literature.}
    \label{tab:VarLit}
    \begin{tabular}{llll}
      \hline \hline
      Reference & \#\ of Quasars & $\Delta T$ Range (yr) & \#\ of epochs \\
      \hline
      \cite{bar94}$^{*}$ & 23 & 0.2$-$1.2 & 2$-$6 \\
      \cite{2007ApJ...656...73L}$^{*}$ & 29 & 0.04$-$0.4 & 2$-$3\\
      \cite{2008ApJ...675..985G}$^{*}$ & 13 & 3.5$-$6.1 & 2\\
      \cite{2010ApJ...713..220G}$^{*}$ & 14 & 0.04$-$6.8 & 2$-$4 \\
      \cite{2011MNRAS.413..908C}$^{*A}$ & 24 & 0.02$-$8.7 & 2$-$13 \\
      \cite{HA12} & 17 & 0.001$-$0.9 & 6 \\
      \cite{2012ApJ...757..114F} & 19 & 1.1$-$3.9 & 2$-$4 \\ 
      \cite{2013ApJ...777..168F}$^{*}$ & 291 & 0.0006$-$3.7 & 2$-$12 \\
      \cite{2015ApJ...806..111G}$^{*}$ & 1 & 0.003$-$0.3376& 32 \\
      \cite{HB15} & 188 & 0.001$-$3 & 2 \\
      This work & 105 & 0.005$-$3.31 & 3$-$7 \\
      \hline
    \end{tabular}
    \begin{tablenotes}
      \item[A]{These quasars are the same (plus one) as those
      in \citet{bar94}; they were re-observed for a longer time baseline between
      observations.}
      \item{$^*$ Data was taken from these works to help create Figure~\ref{fig:deltaEW}.}
    \end{tablenotes}
  \end{threeparttable}
\end{table}

The cause of BAL-trough variability, emergence, and disappearance is still largely
debated in the literature. However, it is likely due to transverse
motion of absorbing clouds across our line of sight (e.g., \citealt{fbqs1408}),
to changes in the ionization of the absorbing gas
(e.g., \citealt{2008MNRAS.391L..39H},
\citealt{2013ApJ...777..168F}, \citealt{2013ApJ...775...14R}),
or to a mixture of these two scenarios.
Multiple spectral observations of BAL variability
can potentially determine how often each of the above causes is at work, which
could significantly increase our understanding of both the physics of the quasar's
outflows and the interaction of the quasar with its host galaxy.
By tracking an absorption feature's emergence, variability, and disappearance may lead
to better predictive power of what BALs may do in the future.

In this work, we analyze the emergence of
broad absorption in quasars. We define emergence to be any BAL that was
previously not present in a quasar but appeared in a newer observation
(this can occur both in BAL and non-BAL quasars).
In contrast to previous works, however, we were motivated to study how
emergent BAL troughs act in future observations.
Does the variability of a trough between two
observations predict or otherwise inform how the trough will vary in a third
observation? We were further interested in how multiple troughs from the same ion
in a single target behave: do troughs vary independently or in a coordinated
fashion? In this study we track the variability of emergent broad absorption
troughs in 105 quasars over at least 3 epochs and up to as many as 7 epochs
per target.

This work is organized as follows.
In \S~\ref{sec:data} we explain
where the quasar dataset came from, how it was selected, and the follow-up
observations we performed to reach 3 or more epochs per quasar.
In \S~\ref{sec:method} we explain our methodology and characterize the BALs in our dataset.
In \S~\ref{sec:discuss} we analyze and discuss the nature of the variability we observe in our dataset and offer possible physical explanations for it.
In \S~\ref{sec:conclude} we summarize our results.

In this work, we at times plot BAL troughs in velocity space using
\begin{equation}\label{e_beta}
\beta\equiv \frac{v}{c}=\frac{R^2-1}{R^2+1} {\rm ~~where~~} R\equiv \frac{1+z_{em}}{1+z_{abs}},
\end{equation}
where $z_{em}$ is the redshift of the quasar, and $z_{abs}$ is the redshift of
the absorbing gas (\citealt{fol86}, \citealt{sdss123}).
We define the zero velocity for each line using its
laboratory vacuum rest wavelength. Throughout this work, we use the
shorter-wavelength member of the \civ\ doublet, 1548.202~\AA,
to define the zero velocity for all \civ\ BALs.
Where needed, we adopt a flat cosmology with
$H_0=70$~km~s$^{-1}$~Mpc$^{-1}$, $\Omega_M=0.3$, and $\Omega_\Lambda=0.7$.

\section{Data}\label{sec:data}

The dataset of emergent BAL troughs analyzed in this work was determined in a
two-step data collection process. First, a candidate sample of emergent broad
absorption lines in quasars was found by visual comparison of older spectra
to newer spectra in publicly available archival data.
Second, a subset of the candidate sample determined by the visual inspection
was re-observed by applying for observing time on the twin Gemini Observatories.
This resulted in at least three spectral observations for a large number of
quasars with emerging broad absorption.

\subsection{Target Selection}\label{sec:targetSelection}

The Sloan Digital Sky Survey (SDSS; \citealt{yor00}) operated
from 2000$-$2008 as SDSS-I and II (hereafter
referred to as SDSS). The SDSS used a dedicated
2.5 meter f/5 Ritchey-Chr\'{e}tien altitude-azimuth telescope located at
Apache Point Observatory in New Mexico, USA \citep{gun06}. The telescope was
outfitted with a photometric camera, detailed in \citet{gun98},
and a multi-object, fiber-fed spectrograph with a wavelength
coverage of 3800$-$9200~\AA\ and a resolving power from 1500$-$3000
(see \S~2 of \citealt{bosssmee}). During its operations, the SDSS collected over
1.5 million spectra of galaxies, quasars, and stars over an area of approximately
10,000~deg$^2$ on the sky. The full catalog can be found in the SDSS Data Release Seven (DR7; \citealt{dr7}), which was publicly available as of 2009.
The DR7 quasar catalog contained 105,783 spectroscopically confirmed quasars \citep{dr7q}.

After SDSS-II concluded, the telescope was upgraded for a third iteration,
SDSS-III \citep{sdss3}.
SDSS-III operated from 2008$-$2014, was a dedicated spectroscopic project,
and executed four different surveys including
the Baryon Oscillation Spectroscopic Survey (BOSS; \citealt{bossover}).
For use in the BOSS, a new multi-object fiber-fed spectrograph was built with
greater throughput, an increased wavelength coverage (3560$-$10400~\AA),
and similar resolving power as the original SDSS spectrograph
(see \S~3 of \citealt{bosssmee}).

On 31 July 2012, the SDSS-III collaboration made public
the SDSS Ninth Data Release (DR9), which included data taken from December
2009 to July 2011 \citep{dr9}. It expanded the original sky coverage
of SDSS to approximately 15,000~deg$^2$ and included spectra for thousands
of quasars not previously targeted spectroscopically by SDSS.
The DR9 quasar catalog was released simultaneously through
the SDSS website\footnote{https://www.sdss3.org/dr9/algorithms/qso\_catalog.php}
and is described in \citet{bossdr9q}.
The DR9 quasar catalog contains 87,822 quasars.

On 29 July 2013, the SDSS-III collaboration made public the SDSS Tenth
Data Release (DR10), which added data taken over December 2009 to
July 2012 \citep{dr10}.
The DR10 quasar catalog was released simultaneously through the SDSS website\footnote{https://www.sdss3.org/dr10/algorithms/qso\_catalog.php}
and is described in \citet{bossdr10q}.
The DR10 quasar catalog contains 166,583 quasars.


We leveraged the multi-epoch nature of the DR7, DR9, and DR10 catalogs to search
for a set of quasars that had been spectroscopically observed at least twice
over all three data releases.
We restricted the search to quasars at $z>1.68$ to be able to search for
\civ\ absorption out to a blueshifted velocity of at least 25,000 \kms.
The search was done using the online SDSS-III CasJobs SQL
tool\footnote{http://skyserver.sdss.org/casjobs/}
by matching the DR7 right ascension (RA) and declination (dec) to the DR9 and
DR10 values to within 2 arcsec. This tolerance is required because the
respective data reduction and astrometric calibration pipelines of all three
data releases produce small differences in on-sky coordinates.

There were 8317 quasars at $z>1.68$ matched between DR7 and DR9.
We refer to this as the DR7$-$DR9
BAL emergence
parent sample.
For each unique quasar, both the DR7 and DR9
spectra were visually compared by plotting both DR7 and DR9 spectra over
top of each other centred on the region between rest-frame $1200-1700$~\AA,
where we expect to find \civ\ broad absorption.
Visual inspection of all parent sample quasar spectra yielded 111 candidates
for the emergence of BAL troughs.
(76 of them were visually classified as BAL quasars within which a second  trough appeared, and 35 were visually classified as non-BAL quasars in DR7.)

Excluding objects previously matched between DR7 and DR9,
there were 8239 quasars at $z>1.68$ matched between DR7 and DR10.
We refer to this as the DR7$-$DR10
BAL emergence
parent sample.
Visual inspection of these objects following the same procedure used for the
DR7-DR9
BAL emergence
parent sample yielded 181 candidates possibly exhibiting new BAL troughs
in DR10 (94 were already BAL quasars based on DR7 data, 87 were non-BAL).

There was also a group of 2037 quasars which were not in DR7 but were
discovered in DR9 and were re-observed in DR10; we refer to these objects
as the DR9$-$DR10
BAL emergence
parent sample.
They were also visually inspected and
14
were found to exhibit candidate
emergent BAL troughs.
We included these objects in our emergent absorption sample; the
ones observed with Gemini
are noted in Table~\ref{tab:obs} as objects that have no SDSS1 or SDSS2 observations.
We attribute the small number of emergent BAL candidates in the
DR9$-$DR10
BAL emergence
parent sample
to (1) the short rest-frame time
separations between DR9 and DR10, and (2) many of the targets in DR9
were re-observed in DR10 due to low signal-to-noise ratios
in their initial spectra.

Combining the DR7-DR9, DR7-DR10, and DR9-DR10 samples of quasars with emergent absorption,
in total there were
306
quasars in our search that may be exhibiting the
emergence of broad absorption; this is the candidate emergent absorption sample.
We refer to it as a `candidate' sample because the emergent absorption was only
found through visual detection and not quantitatively identified.
We perform a more rigorous identification in \S~\ref{sec:BI}
and calculate emergence rates and uncertainties in \S~\ref{sec:emergenceRate}.

\begin{figure}[htb]
\centering
\includegraphics[scale=0.55]{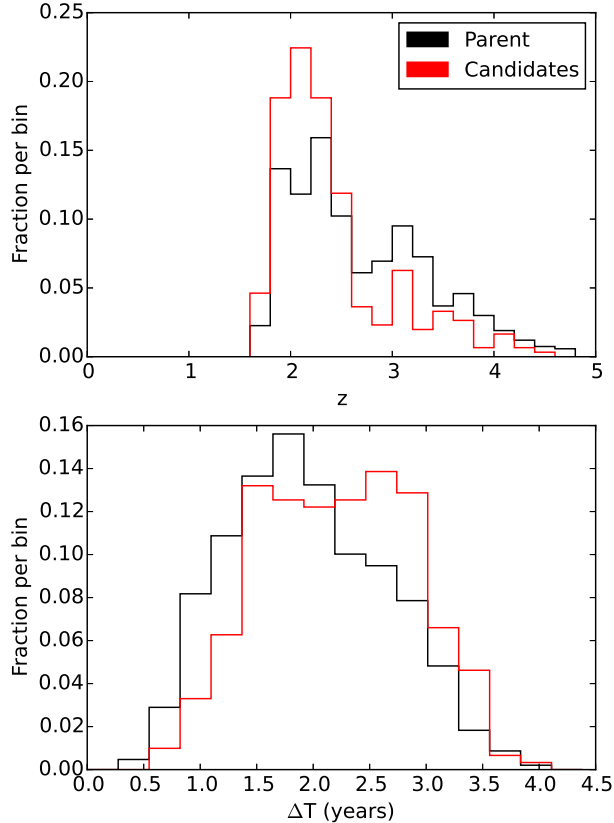}
\caption{The distribution of redshifts (top) and rest-frame time between
the spectra from SDSS and BOSS (bottom) for the parent sample (black) and
the visually determined candidate emergent absorbers (red).}
\label{fig:parentSample}
\end{figure}

In the top portion of Figure~\ref{fig:parentSample} the distributions of
redshifts are plotted for both the parent sample (black) and the candidate
emergent absorption sample (red). In the bottom portion of Figure~\ref{fig:parentSample}, the
distributions of time between observations, $\Delta T$, have been plotted. Again,
the parent sample is in black and the candidate emergent absorption sample is
in red. The candidates are preferentially at smaller
redshifts and the candidates have a bias toward longer time separation between
observations than the parent sample from which they were chosen.
The redshift bias is understandable since higher-redshift spectra typically
have a lower signal-to-noise ratio, making the detection of emergent absorption
more difficult.
The time separation bias arises because BAL quasar troughs tend to show
greater variability on longer timescales (see \S 1).

\subsection{Gemini Observations} \label{gemini}

We obtained data on 105 targets from the candidate emergence sample using the
twin Gemini telescopes.
To select these 105 from the
306
visually identified to have emergent BAL activity,
we prioritized targets which were candidates for non-BAL to BAL quasar transitions, targets with candidate \ion{C}{4} absorption at $v>30,000$ km s$^{-1}$, and brighter targets.
At a given brightness, higher-redshift
targets were preferred so that we could observe the entire 1200$-$1600\,\AA\ continuum
and because of their shorter rest-frame times
between the BOSS epoch and the proposed Gemini observations.

We used the Gemini Multi-Object Spectrograph (GMOS)
(one on each telescope) outfitted with a 1.0\asec\ wide longslit to
observe individually each target in our sample.
For the majority of our observations, we employed the use of the
B600 grating with 600 lines mm$^{-1}$, a blaze wavelength of 461~nm,
simultaneous wavelength coverage of $\sim$300~nm, and $R\sim1688$.
For some high-redshift targets, we used the R400 grating with 400 lines
mm$^{-1}$, a blaze wavelength of 764~nm, a simultaneous wavelength coverage
of $\sim$400~nm, and $R\sim1918$. These settings were
chosen such that the resulting data had similar spectral resolution
to the SDSS/BOSS spectra.  We set our exposure times
to yield a signal-to-noise ratio of $\gtrsim$15 in the rest-frame
1200$-$1600~\AA\ region. This was chosen to be equal to or higher than
the signal-to-noise ratio found in SDSS and BOSS quasar spectra.

Our Gemini follow-up of the 105 targets was spread over three observing
semesters: 2013A, 2013B, and 2014A.
In total, we ran three different observing campaigns, each with multiple
programs on either Gemini North or South.
In Table~\ref{tab:programs} we list
all observing program reference numbers, the number of quasars
observed in that program, and some other notes.
The main campaign was the initial follow-up
observations, targeting all 105 quasars. A smaller, more specific program was
initiated targeting BALs from the main campaign that exhibited variability
in troughs at high-velocity (specifically, BALs blueward of \siv\ emission).
Finally, we were able to utilize a separate Gemini observing campaign
(PhotoVariability) to gather more data on one target.

\begin{table}
\begin{threeparttable}
\caption{A summary of all observing programs on Gemini.
{\bf 
Gemini spectra are available upon request from the authors.
}
}
\label{tab:programs}
\begin{tabular}{lll}
\hline \hline
Program Reference & Objects & Telescope, semester, standard star \\
\hline
Main Campaign$^{A}$ & & \\
\hline
GN$-$2013A$-$Q$-$104 & 14 & North, 2013A, HZ44 \\
GS$-$2013A$-$Q$-$86 & 19 & South, 2013A, LTT6248 \\
GN$-$2013B$-$Q$-$59 & 15 & North, 2013B, G191B2B \\
GS$-$2013B$-$Q$-$50 & 7 & South, 2013B, LTT7379 \\
GN$-$2014A$-$Q$-$67 & 24 & North, 2014A, HZ44  \\
GS$-$2014A$-$Q$-$24 & 27 & South, 2014A, EG274 \\
\hline
High Velocity Campaign$^{B}$ & & \\
\hline
GN-2014B-Q-75 & 3 & North, 2014B, G191B2B \\
\hline
PhotoVariability Campaign$^{C}$ & & \\
\hline
GN-2013B-Q-39 & 1 & North, 2013B, HZ44 \\
GS-2013B-Q-21 & 1 & South, 2013B, LTT7379 \\
\hline
\end{tabular}
\begin{tablenotes}
  \item[A]{The main observing campaign targeted all 105 targets. There are
  actually a total of 106 numbered above because object J161336 was observed
  both in GS$-$2013A$-$Q$-$86 and GN$-$2014A$-$Q$-$67.}
  \item[B]{The high-velocity campaign gathered multiple spectra of 3 targets
  that were already observed in the main campaign: J073232, J083017, J083546.}
  \item[C]{The aim of this spectroscopic followup campaign (see \citealt{2012ASPC..460..124R})
  was only tangentially related to the science of this work; however, we were able to
  use the Standard Target of Opportunity feature of Gemini to observe J023011 twice.}
\end{tablenotes}
\end{threeparttable}
\end{table}

\begin{figure}[h]
\centering
\includegraphics[scale=0.55]{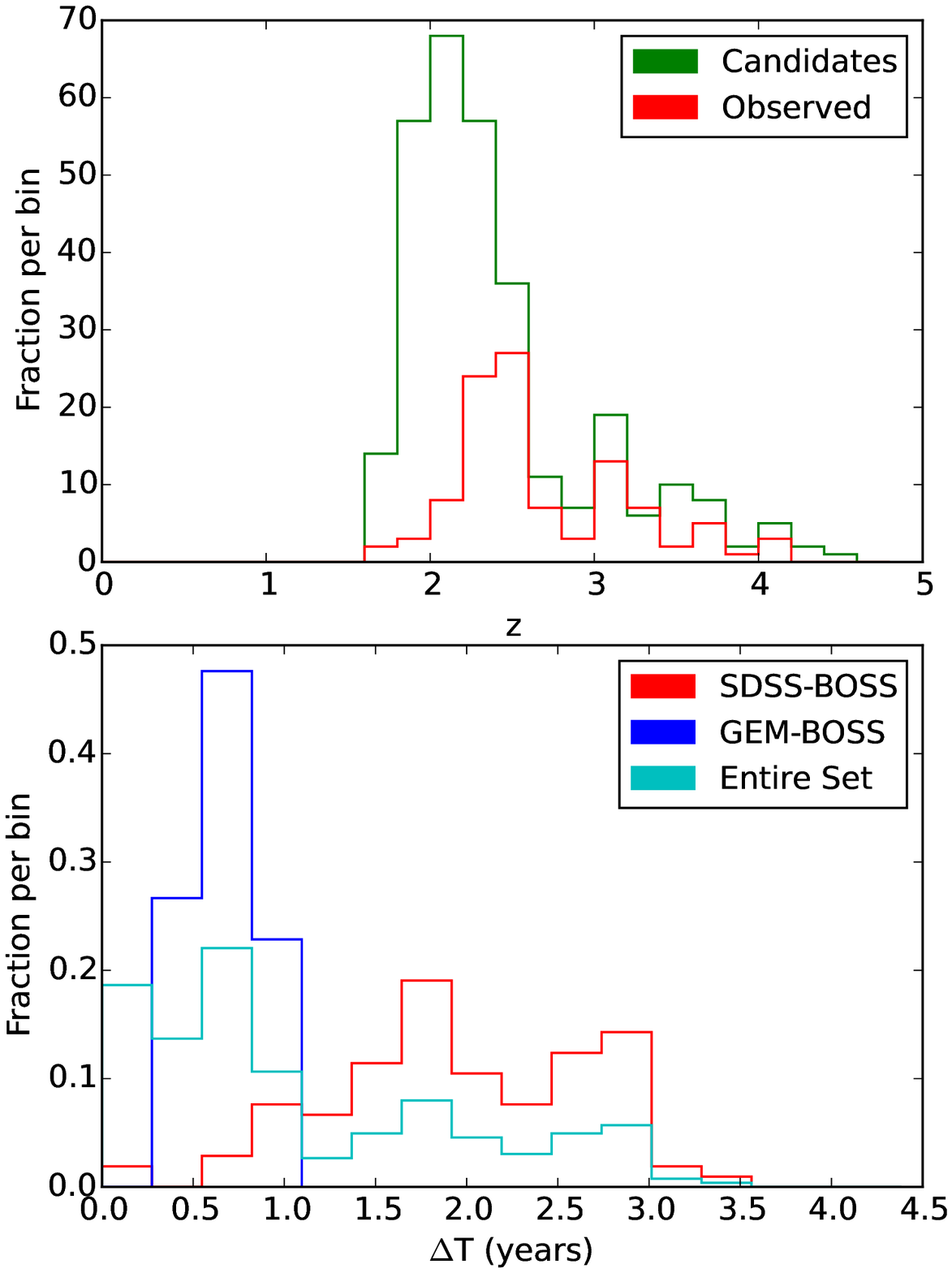}
\caption{\textbf{Top:} The redshift distribution of
the visually determined candidate sample (green), and the subset of those that
received follow-up Gemini observations (red).
\textbf{Bottom:} Rest-frame time separations between {\sl successive} observations
of targets that received follow-up observations on Gemini.
The SDSS-BOSS histogram shows the successive-observation time separations of the last SDSS observation and the first BOSS observation of each quasar.
The GEM-BOSS histogram shows the successive-observation time separations of the last BOSS observation and the first Gemini observation of each quasar.
The Entire Set represents all successive-observation time separations in our dataset, including those involving additional archival epochs or Gemini observations.
See Table~\ref{tab:obs} for further clarification.}
\label{fig:geminiSample}
\end{figure}

In the top portion of Figure~\ref{fig:geminiSample}, the distribution of
redshift is plotted for the original visually determined candidate sample
(green) and those that were observed on Gemini (red).
In the bottom portion of this plot the distributions of time between
observations have been plotted, $\Delta T$.
The red histogram is the $\Delta T$ values for
the objects traced out by the red histogram in the top plot. The blue
histogram is the rest-frame time between the most recent BOSS observation, and
our Gemini follow-up observations.
The cyan histogram represents a number of data
sets, as follows: while our target selection was predicated on just one epoch
from each of DR7, DR9, and DR10, there are many cases where
there are multiple epochs from each data release for each target. Further,
a small subset of our targets from the Main Campaign were observed again
in other campaigns. The cyan histogram captures all rest-frame time between
all {\sl successive} spectral epochs available for a given target from DR7, DR9, DR10,
and Gemini.
See \S~\ref{sec:obsTable} for a full summary of our observational data.

\subsection{Gemini Reductions}\label{redux}

The Gemini spectra were reduced using the Gemini
IRAF\footnote{https://www.gemini.edu/node/10795}
package created by the observatory, following standard techniques.


Only one set of standard star spectra was taken
over the course of each observing program (one for each GMOS wavelength
setting used in the program). Each star is listed
in Table~\ref{tab:programs} and was taken from \citet{lan92}.
Due to the queue nature of our observations, this meant the calibration stars were not
measured on the same night as the quasars themselves and could be separated in
time by as much as 5 months. Thus, while we can use the standard stars to
correct the shape of the quasar spectrum, they cannot be spectrophotometrically
calibrated.


All SDSS and BOSS spectra have wavelength scales based on vacuum
wavelengths.\footnote{https://www.sdss3.org/dr9/spectro/spectro\_basics.php}
The Gemini wavelength calibrations are based on atomic transitions measured
in air. In order to properly compare all data in this work, it was necessary
to shift the Gemini spectra wavelength scale into the vacuum frame. The standard
for this conversion is given in equation (3) of \citet{mor91}.

\subsection{Normalizing the Spectra}\label{sec:normalization}

Our data consist of at least three spectra from three different instruments
attached to multiple telescopes. Moreover, for each quasar, observations could
be separated by as much as a decade in the observed-frame.
To compare easily the relative strengths of the absorption features in data from different telescopes and instruments, we normalized the spectra via the following approach.

In the rest-frame, the ultraviolet-optical continuum of a typical quasar can
be modeled as a power law (to an acceptable approximation). Assuming all quasars are
represented by this model, a power law can then be fit to the continuum
of each quasar and divided out. The resulting normalized spectrum would then
have the continuum resting at a unit-less normalized flux density of 1.0, any
emission features would be greater than 1.0, and any absorption features would
be less than 1.0.

To fit a power-law to the continuum, we chose normalization windows on the
spectra that were relatively free of emission or absorption thus making these
regions a clean sample of the continuum.
Normalization windows for all 105 targets were chosen by eye with the goals of
avoiding all absorption features and as many emission features as possible
and of selecting regions available in all spectra of each target.
In \citet{2012ApJ...757..114F}, six relatively line-free regions
were identified and used for normalization purposes:
1250$-$1350, 1700$-$1800, 1950$-$2200, 2650$-$2710, 2950$-$3700, and
3950$-$4050~\AA. These regions were also identified in \citet{sdss73} as
being relatively clear of any emission features.
Unfortunately, the Gemini spectra have much narrower wavelength coverage than the SDSS and
BOSS spectra; hence, the Gemini spectra narrowed our selection of normalization windows.
Most of the Gemini spectra extend only to $\sim$1750~\AA.

It is important that all chosen normalization windows are the same for all epochs in a given quasar, so that we can be sure the continuum has been fit consistently in all data.
As a result of this requirement, we are unable to use most of the line-free
regions from other works mentioned above. A further result of the smaller
wavelength coverage of the Gemini data is that we were forced to use
normalization windows in the region where we are searching for absorption
(i.e., between 1200$-$1550~\AA), though as much as
possible we chose normalization windows from the list above.

\subsection{Tabular Summary of Observations}\label{sec:obsTable}

The complete list of 105 quasars observed
in the Gemini campaigns is provided online in machine-readable format.
In Table~\ref{tab:obs}, we have provided a portion of the list to guide the reader.
Each quasar is labeled by its SDSS DR7 name in the first column,
though in this work we typically refer to each object by just the first
half of that name. In the second column is the quasars redshift. The modified julian
day (MJD) of the observations
we collected from SDSS, BOSS, and Gemini are labeled in the columns thereafter.
In our dataset, each quasar was observed at least three times, once from each of
SDSS, BOSS, and Gemini. Many quasars were observed more often than that, with as many
as two observations coming from SDSS, two coming from BOSS, and five from the Gemini
observations, making a total of nine possible observations. In Table~\ref{tab:obs},
if we have left a `-' marking a position where no observations were made.
In Table~\ref{tab:obsDeltaT}, the rest-frame time measured in days between successive
observations for a given target is given. As for Table~\ref{tab:obs}, only a portion
of the table is provided, with the entire table availble online in machine-readable format.


\begin{table}
    \caption{A list of the observations made for each of the 105 quasars in this work. The
    objects designation from SDSS DR7 is in the first column, and the MJDs of each
    observation are labeled thereafter. There were a maximum of two observations from SDSS,
    two from BOSS, and five from our Gemini campaigns, though not all objects had that
    many observations.}
    \label{tab:obs}
    \begin{tabular}{lllllllllll}
      \hline \hline
      DR7 Designation & z & SDSS1 & SDSS2 & BOSS1 & BOSS2 & GEM1 & GEM2 & GEM3 & GEM4 & GEM5 \\
      \hline
022143.19$-$001803.8 & 2.65 & 51869.26 & 54081.18 & 55477.34 & -        & 56654.10 & - & - & - & - \\
022559.78$-$073938.8 & 3.02 & 51906.31 & -        & 55832.86 & -        & 56604.27 & -        & -         & - & - \\
023011.28$+$005913.6 & 2.47 & 52200.39 & 52942.34 & 55208.64 & 55454.94 & 56519.53 & 56649.21 & 56685.07 & - & - \\
073232.80$+$435500.4 & 3.46 & 53314.44 & -        & 55180.31 & -        & 56567.61 & 56924.60 & 56951.56 & 57006.41 & 57032.30 \\
074711.14$+$273903.3 & 4.13 & 52592.49 & 52618.32 & 55536.28 & -        & 56569.60 & - & - & - & - \\
081114.66$+$172057.4 & 2.33 & 53357.32 & -        & 55579.23 & -        & 56327.49 & - & - & - & - \\
081811.49$+$053713.9 & 2.52 & 52737.48 & 52962.51 & 55888.42 & -        & 56594.60 & - & - & - & - \\
082313.06$+$535024.0 & 2.56 & 53299.48 & 53381.75 & 55181.90 & 56748.12 & 56328.36 & - & - & - & - \\
082801.67$+$411937.2 & 2.55 & 52265.88 & 54524.13 & 55513.42 & -        & 56327.52 & - & - & - & - \\
083017.31$+$413521.5 & 2.21 & 52265.88 & 54524.13 & 55513.42 & -        & 56335.27 & 56956.58 & 57013.44 & 57046.33 & - \\
      \hline
    \end{tabular}
    \tablecomments{Table \ref{tab:obs} is published in its entirety in the machine-readable format.
      A portion is shown here for guidance regarding its form and content.}
\end{table}

\begin{table}
    \caption{The rest-frame time between successive observations of a given quasar.}
    \label{tab:obsDeltaT}
    \begin{tabular}{lllllllll}
      \hline \hline
      DR7 Designation & $\Delta$12 & $\Delta$23 & $\Delta$34 & $\Delta$45 & $\Delta$56 & $\Delta$67 & $\Delta$78 & $\Delta$89 \\
      \hline
022143.19$-$001803.8 & 606.74 & 382.97 & 322.79 & -      & -     & -     & - & - \\
022559.78$-$073938.8 & 975.73 & 191.69 & -      & -      & -     & -     & - & - \\
023011.28$+$005913.6 & 213.63 & 652.55 & 70.92  & 306.53 & 37.34 & 10.32 & - & - \\
073232.80$+$435500.4 & 418.18 & 310.93 & 80.01  & 6.04   & 12.29 & 5.8   & - & - \\
074711.14$+$273903.3 &   5.04 & 568.99 & 201.5  & -      & -     & -     & - & - \\
081114.66$+$172057.4 & 667.38 & 224.75 & -      & -      & -     & -     & - & - \\
081811.49$+$053713.9 &  63.87 & 830.49 & 200.44 & -      & -     & -     & - & - \\
082313.06$+$535024.0 &  23.13 & 506.06 & 322.3  & 118.01 & -     & -     & - & - \\
082801.67$+$411937.2 & 636.10 & 278.66 & 229.31 & -      & -     & -     & - & - \\
083017.31$+$413521.5 & 703.39 & 308.14 & 255.98 & 193.52 & 17.71 & 10.25 & - & - \\
      \hline
    \end{tabular}
    \tablecomments{Table \ref{tab:obsDeltaT} is published in its entirety in the machine-readable format.
      A portion is shown here for guidance regarding its form and content.}
\end{table}



\begin{figure}
\plotone{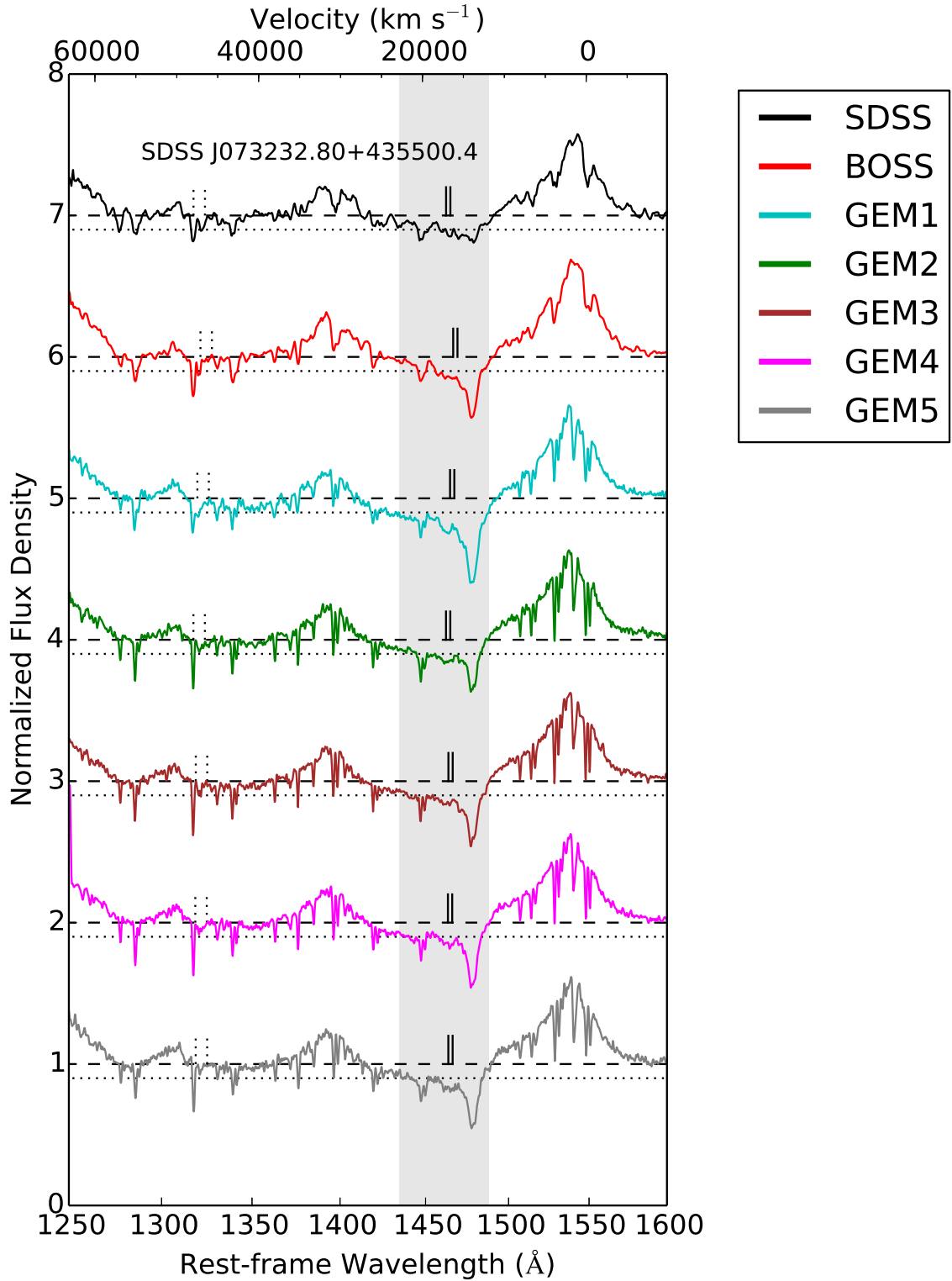}
\label{fig:bigset}
\caption{All spectra gathered for SDSS J073232.80$+$435500.4 are plotted in this figure.
The rest-frame wavelength in Angstroms (\AA) is on the bottom x-axis. The top x-axis
displays the outflow velocity in \kms\ relative to the systemic redshift of the quasar,
as measured from the \civ\ emission feature at ∼1550~\AA. The y-axis is normalized Flux
Density units. Each spectrum available for this quasar is plotted, artificially
separated in the y-direction to make comparisons between epochs easier.
For each spectrum, there is a dashed line placed at the normalized continuum level,
and a dotted line placed at 90\% of the normalized continuum level.
The spectrum with the smallest MJD is at the top, and each successive epoch in
chronological order is placed below. The colors represent different data, which is
given in the legend. In each spectrum, the centroid velocity of each absorption complex
is denoted by two small vertical black bars, which represent the \civ\ doublet at
\lala~1548.202,1550.774~\AA. The expected locations of \siv\ absorption that could
accompany each \civ\ doublet are denoted by two small vertical dashed lines,
which represent the \siv\ doublet at \lala~1393.755,1402.770~\AA.
This is an example plot. A plot for all 105 quasars in this study is available
in the electronic version of this publication.}
\end{figure}

\subsection{Signal-to-Noise and Smoothing}\label{sec:SNR}

Our data show a range of SNR for each spectral epoch due to the observations' different telescopes, instruments, and observing conditions.
To compare these spectra properly, we applied the following smoothing routine.
We measure SN$_{1500}$, the median (flux density)/(1$\sigma$ noise)
value for all pixels at $1400-1600$~\AA,
and determine which spectrum has the highest SN$_{1500}$ for the quasar.
Each other spectrum of this quasar is smoothed with a boxcar filter an odd number
of pixels wide, using the smallest number that yields SN$_{1500}$ at least equal to highest SN$_{1500}$ found for the quasar.  To avoid degrading the effective resolution
of the spectra, we capped the boxcar width at 9 pixels.

\subsection{Identification of Broad Absorption Lines}\label{sec:BI}

The targets selected in \S~\ref{sec:targetSelection} were found by visual
identification of emergent broad absorption (see \S~\ref{sec:targetSelection}
for details). This was a satisfactory approach
to determine an initial set of quasars to study but did not include a formal
definition of absorption that confirmed the visual emergence as statistically
significant absorption.
A formal definition that would cleanly separate quasars into those with
broad absorption lines (BAL quasars) and those without (non-BAL quasars) is
required in order to reduce subjectivity
and allow comparison of the trough strength in BAL quasars.

In response to this need, \citet{wea91} defined the BALnicity Index (BI).
The original definition of the BI measured the amount of
absorption (in units of \kms) of \civ\ by integrating over all available
absorption troughs in a spectrum
while applying the following criteria.
Any absorption that is included in the index must
be at least 2,000~\kms\ wide;
any absorption found within
the first 3,000~\kms\ blueward of the \civ\ emission peak (at $\sim$1550~\AA)
is ignored;
no absorption after
25,000~\kms\ blueward of the \civ\ emission peak is included;
and, the flux
density in the trough must be below 90~\%\ of the normalized flux density at the
continuum.


As research into BAL quasars progressed, it was realized that this
original definition of BALnicity erred on the conservative side; it does not
fully encompass all intrinsic broad absorption observed in quasar spectra.
For example, a trough narrower than the 2,000~\kms\ criterion can still be
rooted in the same physical origin as broader troughs and should be included.
These are known as `mini-BAL' troughs in the literature and range in width from
500$-$2,000~\kms\ (e.g., \citealt{2011MNRAS.411..247R}).
Modifications of the BALnicity Index have been proposed (e.g., \citealt{sdss123}; \citealt{trump06}) to include absorption over wider velocity ranges.

For this work, in which we are interested in troughs which may be narrower or at
higher outflow velocities (or both) than considered in the traditional BI
definition, we measure BALnicity using a modification of the index
proposed in \citet{trump06}, which we define as:
\begin{equation}
{\rm BI}^* = \int_{v_{low}}^{v_{high}}\left ( 1 - \frac{f(v)}{0.9} \right ) C_{1000} ~dv,
\label{eq:BI}
\end{equation}
using the asterisk to distinguish it from the original BI definition.
In the above, $f(v)$ is the normalized flux density at a given velocity and
the quantity $C_{1000}$ is equal to 1 only in regions more than 1,000~\kms\ wide in which
the quantity in parentheses is everywhere greater than zero, otherwise
it is set to 0.  That change is similar to \citet{trump06}.
In the above equation, $v_{low}$
and $v_{high}$ are purposefully not defined to indicate there is no formal limit
on the minimum and maximum absorbing velocities. (A minimum velocity of zero would
exclude the rare but interesting redshifted-trough BAL quasars; see \citealt{rsbal1}).
In \S~\ref{sec:targetSelection}, targets with possible absorption anywhere
between the \lya\ emission and \civ\ emission were chosen. Historically,
the region betwen \lya\ and \siv\ emission lines was excluded because this
region could be contaminated by \siv\ absorption associated with \civ\
absorption. This is why the original BALnicity measurements were capped
at 25,000~\kms\ blueward of the \civ\ emission peak. We specifically
searched for absorption from \civ\ in both the classically searched velocity
regimes, but also in much higher velocities regimes. Thus, while no
formal limits on the BALnicity were required for this work, in practice the
bounds on \BI\ were set to $0 < v~(\textrm{\kms}) < $ 65,000.
If \BI$>0$ we consider there to be statistically
significant absorption present; a \BI$=0$ indicates no absorption is present.
After absorption is identified, we then determined if the absorption was
a result of high-velocity \civ\ or accompanying \siv\ (see Fig.~\ref{fig:bigset}, and the next section,
for this identification).

\subsection{Measurement of BAL Trough Properties}\label{sec:visinspec}

All normalized spectra were run through an automatic BALnicity measurement
routine. As part of this routine, on top of the variable smoothing from \S~\ref{sec:SNR} we smoothed by an additional 3-pixel-wide boxcar to implement Savitsky-Golay smoothing, which weights pixels closer to the centre of the smoothing window more than those at the edge of the window (see \citealt{sg64}).

All absorption meeting the \BI\ requirements identified by the measurement
routine was visually inspected, and any contamination was removed. There were a number
of possible contaminants that required visual confirmation or elimination.
Intervening absorption or narrow \civ\ systems are, by design, meant to be
ignored by \BI. However, narrow systems in spectra that were heavily smoothed by
our technique were sometimes smoothed out enough in velocity to be falsely detected as BAL troughs; such cases were removed by comparing the smoothed
spectrum to the un-smoothed spectrum after the absorption had been identified.
Narrow systems may also occur on-top or blended into a BAL feature.
If this occurred, the contribution of the narrow system was included in
the measurement of \BI. While this means there are some possible contaminations
for the individual \BI\ numbers, it is important to note that this paper
focuses on the difference in absorption from one epoch to another. Assuming
the narrow systems have remained relatively constant over our observation
campaigns, their effects would cancel out in calculating change in absorption.
We also removed \siv\ BAL troughs accompanying
\civ\ troughs. Quasar-specific details regarding how contaminations
were dealt with can be found in that quasar's individual caption in the
online figure set; see Fig.~\ref{fig:bigset}.

After all non-\civ-BAL related detections were removed, there were a total of
653 individual \civ\ absorption troughs across all 360 normalized spectra in the 105 quasars.
To quantify the properties of this sample of absorption troughs,
we measured each trough's centroid velocity (in \kms), width (in \kms), and
fractional depth below the normalized continuum in two ways detailed in the next paragraph.
The centroid velocity, $v_{cent}$, of a trough was measured
following the definition in \citet{2013ApJ...777..168F}: the mean
of the velocity in a trough where each pixel is weighted by its
distance from the normalized continuum.
The width of a trough is the velocity
range over which the trough met the \BI\ criteria.

The mean depth of the trough was calculated in two ways.
First, we measured $d_{BAL}$ as in \citet{2013ApJ...777..168F}, which is the
{\sl mean} depth of the
trough relative to the normalized continuum of 1.0 for each data point in the
trough. Second, we measured $d_{max7}$, a measure of the {\sl maximum}
trough depth which is calculated by sliding a
7-pixel-wide window across the trough, measuring the average depth
over each window relative to the normalized continuum at 1.0, and taking the
largest depth over all these windows as $d_{max7}$. The uncertainty on the
depth is calculated as the uncertainty in the mean of the 7 pixels in the
average. We note that since the observations were taken with different
telescopes and instrument set ups, 7 pixels correspond to slightly different
resolutions; however, the differences do not substantially affect the results.

\begin{figure}[htb]
  \centering
  \includegraphics[width=0.66\textwidth]{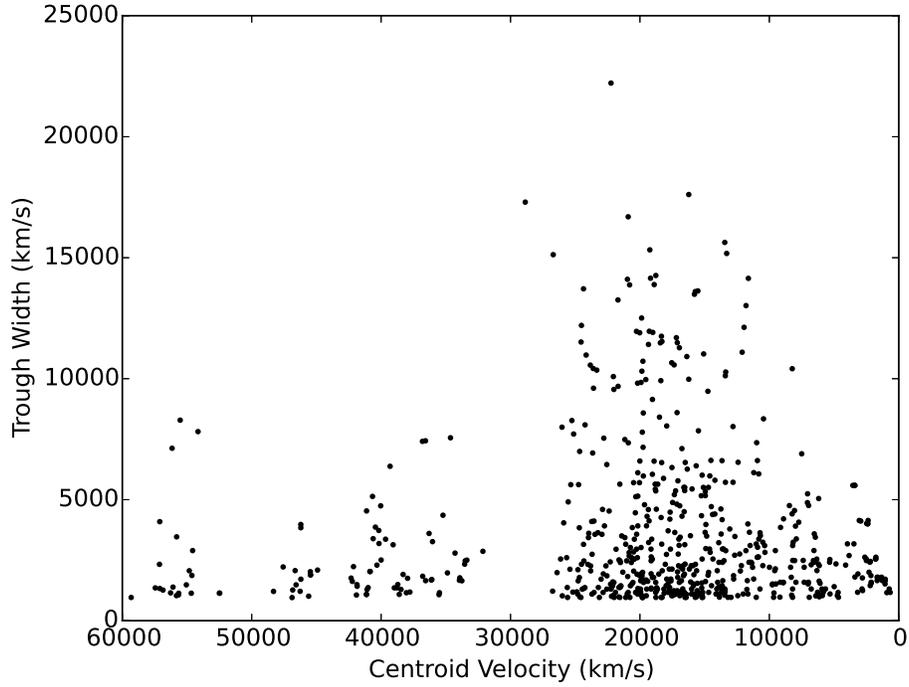}
  \caption{The distribution of trough widths and trough centroid velocities for
  all absorption identified by the \BI\ criteria. There are a total of 653
  absorption features in 360 spectra.
  The gaps near 30,000 and 50,000 \kms\ centroid velocity are due primarily to absorption at wavelengths overlapping \ion{Si}{2} or \ion{Si}{4} emission not reaching the detection threshold of 10~\% below the level of the power-law continuum.
  }
  \label{fig:origcentwidth}
\end{figure}

The distribution of trough width versus trough centroid velocity
is plotted in Figure~\ref{fig:origcentwidth}.
The mean centroid velocity in the plot is 19,500~\kms, and the mean trough
width is 3,600~\kms.
The widest trough, at 22,200~\kms, was observed in J083546; its spectra are plotted in \S~\ref{sec:afteremerge}.
There are noticeable gaps in the centroid velocity distribution at 30,000~\kms\ due
to \siv\ emission at $\sim1400$~\AA, at 43,000~\kms\ due to \cii\ emission at
$\sim1335$~\AA, and at 50,000~\kms\ due to \SIii+\oi\ emission at $\sim1304$~\AA.
Si\,{\sc iv} is on average the strongest of those three lines in terms of emission-line equivalent width and C\,{\sc ii} is the weakest \citep{sdss73}.
Correspondingly, the gap in the distribution of centroid velocities
is widest for \siv\ and narrowest for \cii.
The presence of broad emission either raises the flux density level
from which absorption must remove flux
or adds flux density which does not pass through the absorber on its way to us.
Either case can make it more difficult for the deepest part of the trough from a
given absorber to drop below our detection threshold of 0.9 times the continuum level.
Thus, while there is likely absorption
present at these velocities, only the strongest absorption is detected.

\begin{figure}[htb]
  \centering
  \includegraphics[width=0.66\textwidth]{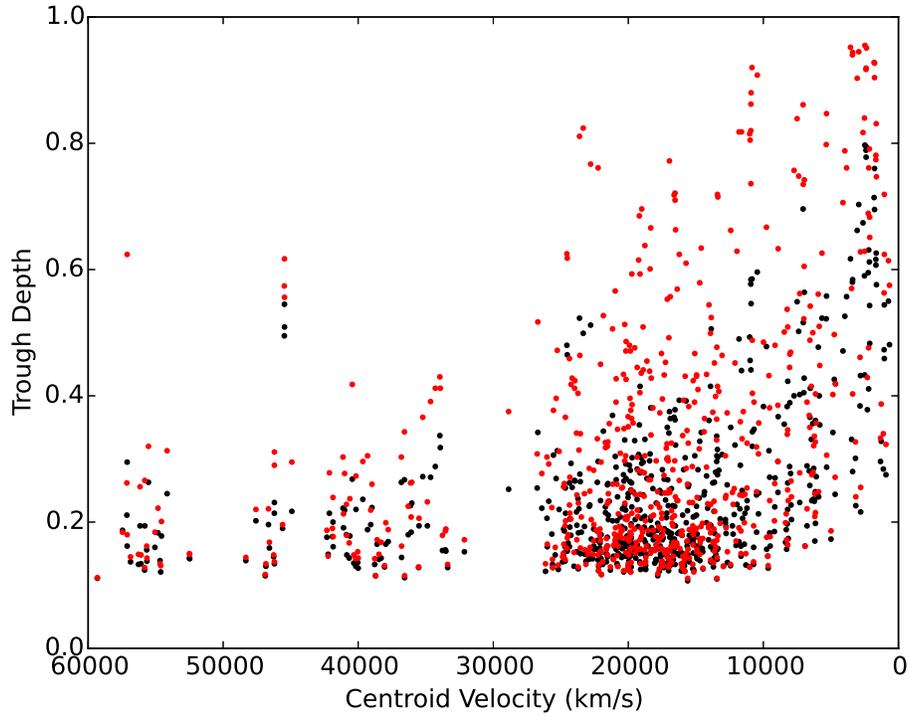}
  \caption{Comparing the depth of troughs relative to the normalized continuum at 1.0,
  versus their centroid velocities for all absorption identified by the \BI\
  criteria. There are a total of 653 absorption features in 360 spectra.
  The black points are $d_{BAL}$, and the red points are $d_{max7}$.}
  \label{fig:origdepthwidth}
\end{figure}

The distribution of trough depth (both $d_{BAL}$ and $d_{max7}$) versus
trough centroid velocity is plotted in Figure~\ref{fig:origdepthwidth}.
The quantity $d_{max7}$ measures the maximum depth of the trough, and
indeed measures on average larger depths than
$d_{BAL}$, which is representative of the mean depth over the entire trough.
In black is $d_{BAL}$, with a mean
value of 0.24, and in red is $d_{max7}$ with mean 0.32. The $d_{max7}$ value
at $0.6$ and approximately 57,000~\kms\ is from object J105210 and is the
result of an intervening absorber sitting on top of a BAL at high velocity.

\begin{figure}[htb]
  \centering
  \includegraphics[width=0.66\textwidth]{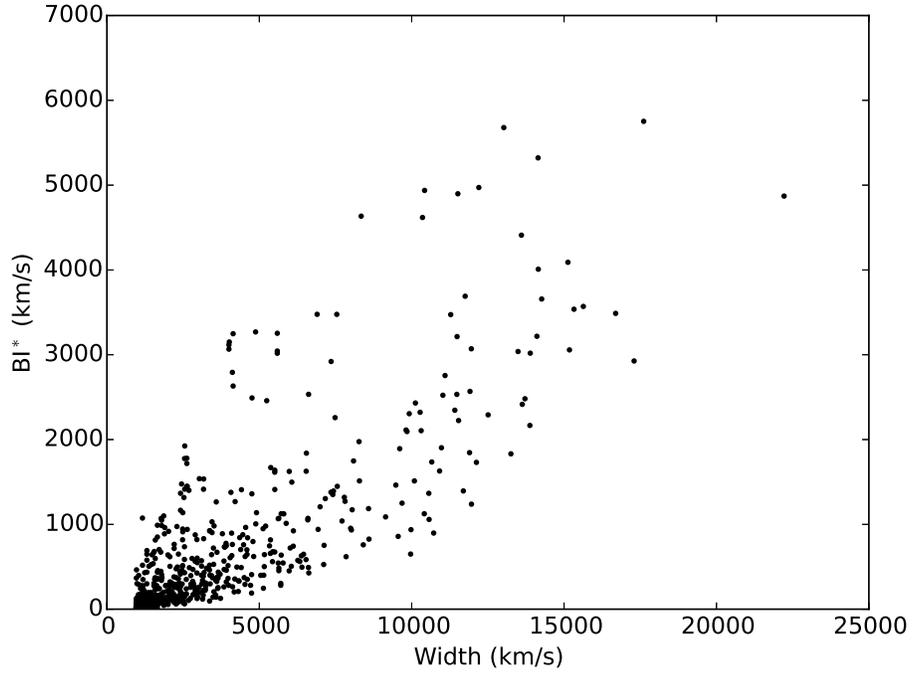}
  \caption{The \BI\ plotted against the width of
  the troughs identified by the \BI\ criteria.}
  \label{fig:origwidthBI}
\end{figure}

In Figure~\ref{fig:origwidthBI}, the trough width is plotted against the
\BI\ of each trough. The mean \BI\ is 678~\kms. The largest value of \BI\
is found in J130600. In its BOSS spectrum, the absorption spans the entire
region between \civ\ and \siv\ emission, reaches $d_{BAL}=0.4$, and width
of 17,600~\kms. It is the second-widest trough, next to J083546 above.

\begin{figure}[htb]
  \centering
  \includegraphics[width=0.66\textwidth]{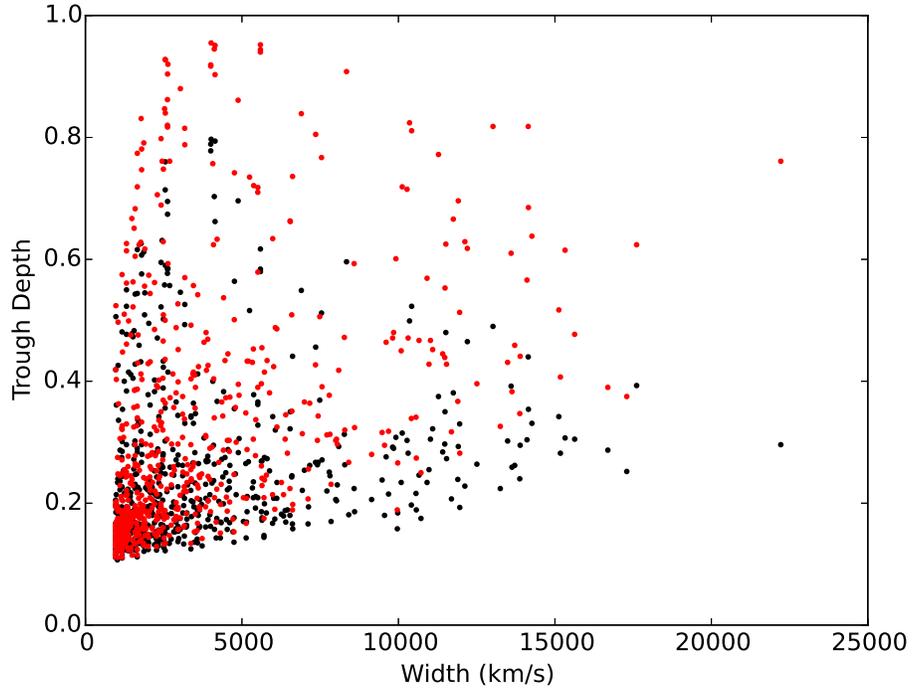}
  \caption{The depth $d_{max7}$ (red) and $d_{BAL}$ (black), relative to the continuum
  at 1.0, plotted against the velocity width of the trough.}
  \label{fig:origwidthdepth}
\end{figure}

Finally, the trough depth (both $d_{BAL}$ and $d_{max7}$) is plotted against
trough width in Figure~\ref{fig:origwidthdepth}. The two data points at very
high trough width are a result of J083546. Again, in this case $d_{max7}$ is
not the best measure of the depth of the trough because it is biased by a
narrow feature sitting on top of the broad absorption.

\clearpage

\clearpage

\section{Methods}\label{sec:method}

\subsection{Absorption Complexes}\label{sec:compident}

\begin{figure}[htb]
\centering
\includegraphics[width=0.75\textwidth]{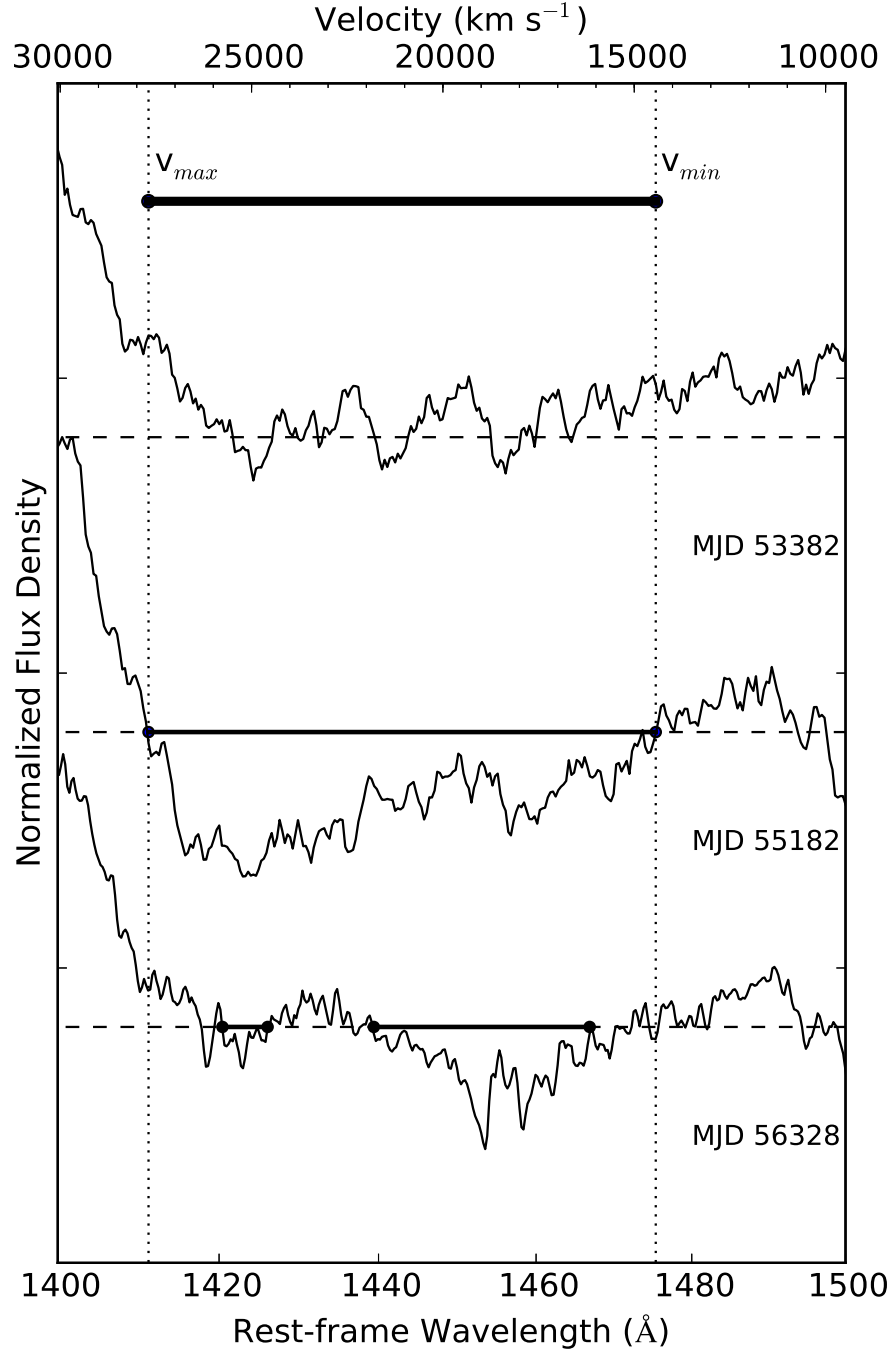}
\caption{
This figure demonstrates an example of how an absorption complex is identified
and measured.
The Spectra is of quasar J082313.06$+$535024. The oldest spectral epoch
is at the top, and each epoch is in chronological order after that (see the
MJD label beside each spectrum).
In all epochs, part of the \ion{Si}{4} emission line is seen just longward of 1400\,\AA.
In the first epoch, there is no absorption,
in the second a broad absorption trough appears, and in the final epoch, that
trough splits into two.
The thin black lines indicate the individual absorption features, and the entire
absorption complex identified by the prescription in \S~\ref{sec:compident} is
labeled at the top from $v_{max}$ to $v_{min}$. Note, that while there are varying
numbers of individual troughs for each epoch, it is considered to be one complex.} 
\label{fig:absComplex}
\end{figure}

From epoch to epoch, a given trough may split into multiple smaller troughs
as it decreases in absorption strength, or two adjacent smaller troughs may
merge into one large trough as the absorption of one or both of the troughs
widens, or the depth increases. In order to characterize properly the
variability of broad absorption between epochs, we identify BAL complexes.
We have mimicked the approach of \citet{2013ApJ...777..168F} in identifying complexes.
In Figure~\ref{fig:absComplex}, we show an example of how this
an absorption complex is identified
for the quasar J082313. In the earliest epoch, no absorption feature
is observed. In the following epoch a large trough emerges,
spanning the region $1411.2-1475.3$~\AA. In the final epoch, that large trough
has split into two separate troughs spanning $1420.4-1426.0$~\AA\
and $1439.4-1466.9$~\AA. We consider this all to be one complex of \civ\
absorption. To identify complexes, we follow these steps:

\begin{enumerate}
  \item{Sort all troughs identified by the BALnicity code above in order from
  highest velocity to lowest. They are also sorted into spectral epoch order,
  from oldest to newest observations.}
  \item{Begin with the highest velocity trough in the oldest spectrum, setting
  the $v_{max}$ and $v_{min}$ of the absorption complex to this trough's
  velocity range values.}
  \item{Loop through all absorption features in the following epochs. If the
  complex's $v_{max}$ intersects a trough in a later epoch, reset $v_{max}$ to
  that trough's $v_{max}$. The same is done for $v_{min}$.}
  \item{This is repeated for all absorption features identified by the
  BALnicity measurements above.}
  \item{Repeat the entire process starting with the most recent epoch
  and moving toward the oldest.  This step ensures that troughs in the
  velocity range of the complex are counted as part of the complex in all epochs.}
\end{enumerate}

As is evident in Figure~\ref{fig:absComplex}, this procedure results in a
maximum and minimum velocity, $v_{max}$ and $v_{min}$, respectively, range that
encompasses all absorption that overlaps in that region for all epochs;
we define the
difference between these two values as the velocity width of the absorption
complex. Thus each of the 653 individual \civ\ absorbers are associated with
one complex. This results in 219 individual absorption
complexes across all 105 quasars in our sample.

\clearpage

\subsection{Quantifying the Absorption}\label{sec:measure}

In order to quantify the variability in the absorption complexes identified
in the previous section, we must measure some of their absorption parameters.
After absorption complexes have been identified in the spectra, we measured
the equivalent width (EW; in \AA, defined below), the weighted centroid
velocity $v_{cent}$, and the average trough depth ($d_{BAL}$ and $d_{max7}$)
over that region for all epochs of a quasar's spectrum. This is regardless of
whether or not absorption is actually identified in the spectrum. For example,
in Figure~\ref{fig:absComplex}, the earliest spectral epoch at MJD~53382
exhibits no absorption, however, we still measure the EW, $v_{cent}$, and depth
of the trough over the complex's range in this epoch. This provides us with a
baseline from which to measure changes.

To measure the EW in \AA\ and its uncertainty from the normalized
spectra we used equations 1 and 2 of \citet{2002ApJ...574..643K}:
\begin{equation}
  \textrm{EW} = \sum_i \left( 1 - \frac{F_i}{F_c} \right) B_i,
  \label{eq:EW}
\end{equation}
and,
\begin{equation}
  \sigma_{\textrm{EW}} = \sqrt{\left[\frac{\Delta F_c}{F_c} \sum_i\left(\frac{B_iF_i}{F_c}\right) \right]^2 + \sum_i\left(\frac{B_i\Delta F_i}{F_c}\right)^2}.
  \label{eq:sigEW}
\end{equation}
$F_i$ and $\Delta F_i$ are the normalized flux density and its error in the
$i$th bin. $F_c$ and $\Delta F_c$ are the mean and the uncertainty on the mean
of the continuum normalized flux density measured in the normalization windows.
$B_i$ is the bin
width in units of \AA. In our
normalized spectra, $F_c=1$ and $\Delta F_c$ values are calculated using the
normalization windows determined by the normalization procedure in
\S~\ref{sec:normalization}.
Thus $\sigma_{\textrm{EW}}$ represents the statistical uncertainty
inherent in spectra. It does not quantify the systemic uncertainty,
which is governed by the placement of the continuum by normalization.
The wavelength range over which the sums in equations (\ref{eq:EW}) and
(\ref{eq:sigEW}) are measured is set by
the identification of $v_{max}$ and $v_{min}$ in the previous
section. The \BI\ is
measured only when the normalized flux density is below 90~\%\ of the total
continuum for more than
1000~\kms. The beginning and ending wavelengths where this criterion is
satisfied are carried over to the absorption complexes, and the EW
is measured between them. Thus, while the range over which equations
(\ref{eq:EW}) and (\ref{eq:sigEW}) are measured is set by this criterion, we
still use $F_c=1$ as the normalized continuum level (and not 0.9 as might
have been expected).

As mentioned in \S~\ref{sec:visinspec}, we use the centroid velocities and
mean depths as tool to compare troughs. We now apply the same measurements
to the velocity range of the absorption complexes with a caveat in the
case of measuring the centroid velocity.
If in the absorption complex, the normalized flux density is above 1.0, the
velocity of that bin is not counted toward the weighted mean $v_{cent}$. If
\textit{all} of the normalized flux density is above 1.0 for the absorption
complex, which can happen in cases where absorption has disappeared on top of
an emission feature, then all bins are weighted equally. This results in a
centroid velocity being in the mean
of the absorption complex's maximum and minimum velocities.
Calculating the mean depths, $d_{BAL}$ and $d_{max7}$, in the absorption
complex is the same as in \S~\ref{sec:visinspec}, i.e.,
the normalized flux density values above 1.0 are not ignored.

\subsection{Measuring Variability}

There are several ways to compare one epoch to the next for any of the
absorption complexes. We adopt several methods including the measurement of change of EW:
\begin{equation}
  \Delta \textrm{EW} = \textrm{EW}_2 - \textrm{EW}_1 \qquad \sigma_{\Delta \textrm{EW}} = \sqrt{\sigma^2_{\textrm{EW}_2} + \sigma^2_{\textrm{EW}_1}}.
  \label{eq:deltaEW}
\end{equation}
We also measure the fractional change in EW, which is the change in EW
from one epoch to the next divided by the average EW over both epochs
(e.g, \citealt{GB08}, \citealt{2013ApJ...777..168F}). This measurement indicates how
significant a change in absorption is compared to the size of the feature that
is changing.
\begin{eqnarray}
    \frac{\Delta \textrm{EW}}{\langle \textrm{EW} \rangle} = \frac{\textrm{EW}_2 - \textrm{EW}_1}{(\textrm{EW}_2+\textrm{EW}_1)\times0.5} \\
    \frac{ \sigma_{\Delta \textrm{EW}} }{\langle \textrm{EW} \rangle } = \frac{4\times(\textrm{EW}_2\sigma_{\textrm{EW}_1} + \textrm{EW}_1\sigma{\textrm{EW}_2})}{(\textrm{EW}_2+\textrm{EW}_1)^2}
  \label{eq:frac}
\end{eqnarray}
Similarly, we measure the change in depth from one epoch to the next, and
those corresponding uncertainties:
\begin{eqnarray}
    \Delta d_{BAL} = d_{BAL,2} - d_{BAL,1} \\
    \sigma_{\Delta d_{BAL}} = \sqrt{\sigma^2_{d_{BAL,2}} + \sigma^2_{d_{BAL,1}}}. \\
    \Delta d_{max7} = d_{max7,2} - d_{max7,1} \\
    \sigma_{\Delta d_{max7}} = \sqrt{\sigma^2_{d_{max7,2}} + \sigma^2_{d_{max7,1}}}.
  \label{eq:deltaDBAL}
\end{eqnarray}

These diagnostics of variability are calculated between all epochs for all
219 absorption complexes in the dataset.

\section{Results and Discussion}\label{sec:discuss}

\subsection{Summary of Absorption Complex Characteristics}

Combining \BI\ requirements with the absorption complex identification
routine resulted in 219 individual \civ\
absorption complexes in 105 quasars. Of the 105 quasars in our dataset,
two quasars, J121314 and J142903, did not end up having any absorption that met the \BI\ criteria for absorption, making the total number of quasars that
contributed data for absorption complexes to this study 103. Both
J121314 and J142903 are still included in the observation Tables~\ref{tab:obs}
and \ref{tab:obsDeltaT}, and the online figure set (Fig~\ref{fig:bigset})
for completeness data was collected for them using Gemini.
There are a total of 354 spectra for
these 103 quasars, as each quasar is observed between 3-7 times (see
Table~\ref{tab:obs}).

\begin{figure}[htb]
  \centering
  \includegraphics*[scale=0.55]{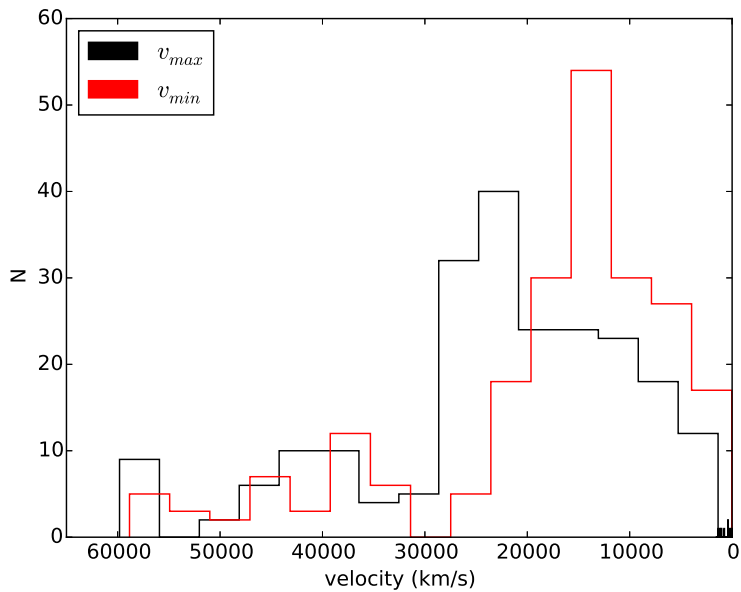}
  \caption{The distribution of absorption complex maximum and minimum velocities,
  $v_{max}$ and $v_{min}$, respectively. The majority of the values are
  between 0 and 30,000~\kms, the region between \civ\ and \siv.
  The distributions nonetheless extend to the largest velocities probed.
  }
  \label{fig:distV}
\end{figure}

The distribution of maximum and minimum velocities, $v_{max}$ and $v_{min}$,
respectively, in the 219 absorption
complexes is plotted in Figure~\ref{fig:distV}. The quasar with the
largest $v_{max}$ was J023011 with 59,800~\kms.
The smallest $v_{max}$ was
found in J091621 at 1,400~\kms,
which also had the smallest $v_{min}$
at just 30~\kms. The largest $v_{min}$ is at 58,800~\kms\
in quasar J113536.

\begin{figure}[htb]
  \centering
  \includegraphics*[scale=0.55]{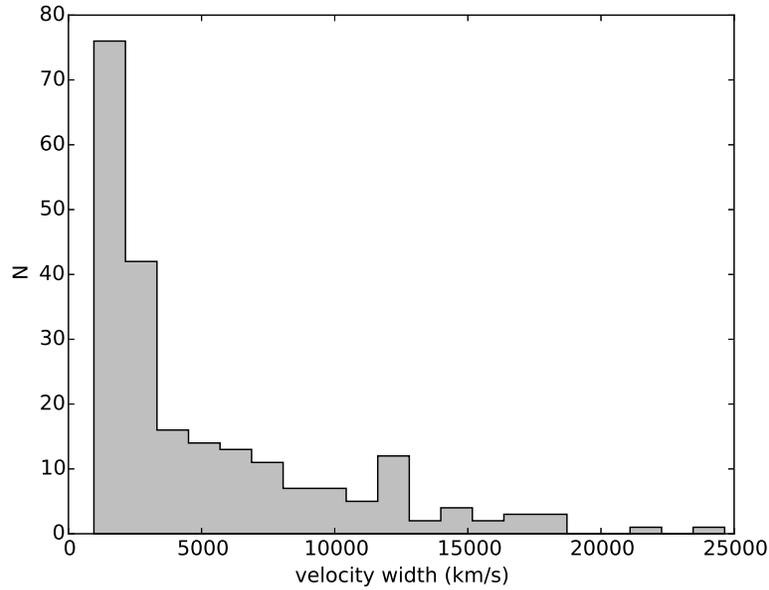}
  \caption{The distribution of absorption complex velocity width. There are 219 data
  points in the histogram. The figure shows that most of the troughs
  have small widths, peaking well below the 5,000~\kms\ value. Note
  that the smallest allowable width was 1,000~\kms.}
  \label{fig:distVwidth}
\end{figure}

We plot the distribution of absorption complex velocity widths, which was
defined in \S~\ref{sec:compident}, in Figure~\ref{fig:distVwidth}.
There are more complexes with smaller widths than with
large widths. The widest absorption complex is found in J165642; it contains
a complex 24,600~\kms\ wide.
The smallest absorption complex width in our sample is 1,000~\kms, which
is the lower limit imposed by the \BI\ criterion. Note that the widths presented
in Figure~\ref{fig:distVwidth} are not the true widths of individual troughs,
but the widths of the complexes.

\begin{figure}[htb]
  \centering
  \includegraphics*[scale=0.55]{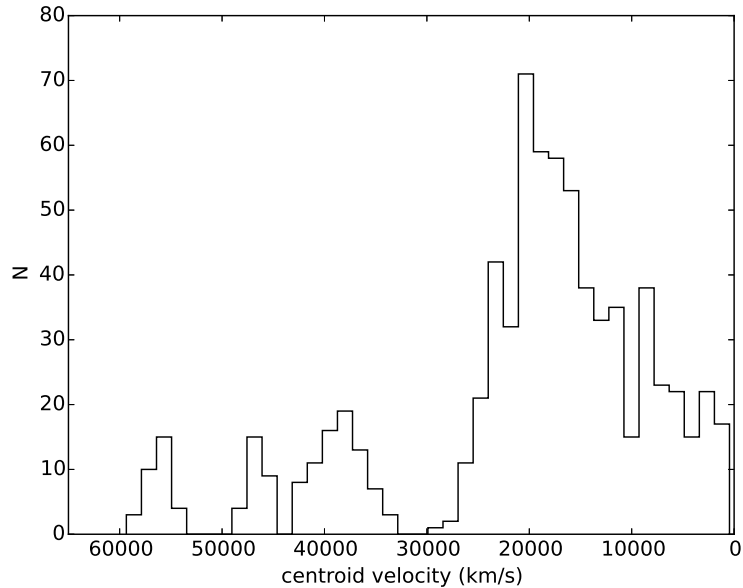}
  \caption{The distribution of absorption complex centroid velocities, $v_{cent}$.
  There are 748 data points in the histogram. The majority of the centroid
  values lie between 0 and 30,000~\kms, the region between \civ\ and \siv.
  Also of note in the figure is the lower number of troughs identified at the
  wavelengths of various emission features: Si\,{\sc ii} at $\sim$50,000~\kms,
  \cii\ at $\sim$45,000~\kms, and \siv\ at $\sim$30,000~\kms.}
  \label{fig:distVcent}
\end{figure}

The distribution of absorption complex centroid velocities, $v_{cent}$, is
plotted in Figure~\ref{fig:distVcent}. There are 219 absorption complexes, each
complex having at least 3 epochs of observations; this results in 748 data
points.

\begin{figure}[htb]
  \centering
  \includegraphics*[scale=0.55]{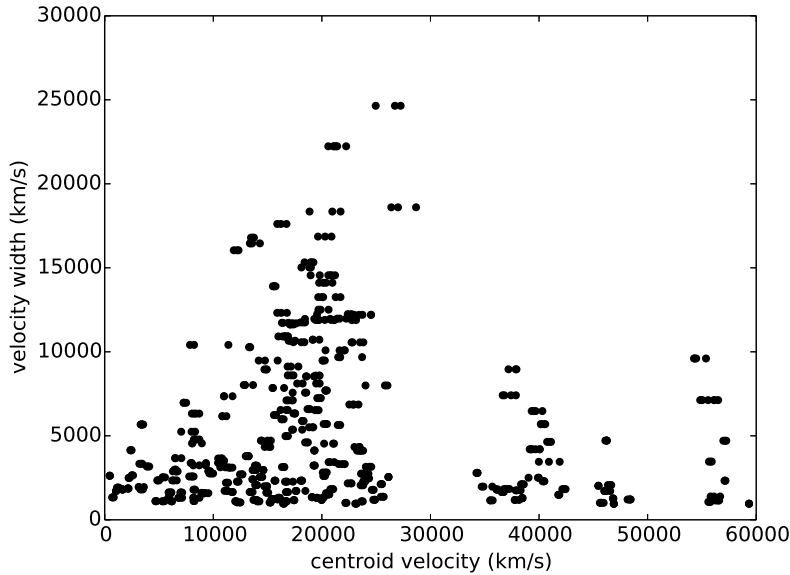}
  \caption{The width of the absorption complexes versus the individual
  centroid velocities measured for each epoch.
  There are 748 data points plotted.}
  \label{fig:centwidth}
\end{figure}

In Figure~\ref{fig:centwidth}, for every complex the centroid velocity
is plotted against the width.
There is a marked absence of troughs wider than 10,000~\kms\ beyond
the wavelength of \siv\ at $\sim$30,000~\kms.

\subsection{BAL/non-BAL Transitions}\label{sec:baltrans}

An emergent BAL quasar is defined as a quasar which is measured to have \BI$=0$
in one spectral epoch and \BI$>0$ in the next spectral epoch.
This change indicates that a significant trough appeared
in a quasar which was previously designated a non-BAL quasar.
Of course, the opposite can happen: all troughs in a BAL quasar can disappear
and change the quasar's designation from BAL to non-BAL.
Such objects were studied extensively in \citet{2012ApJ...757..114F}.

Of the 103 targets with significant absorption in our dataset, there were 36
instances of transition from non-BAL to BAL: emergent BAL quasars.
In all but three of these instances the transition occurred between the SDSS
and BOSS observations, which is expected given that the full sample was chosen
due to visual identification of new absorption between those epochs.
The remaining three transitions occurred in J113536, J145230,
and J222838 between their BOSS and Gemini observations.
In these quasars, emergent absorption is visible in the SDSS-BOSS transition
but it did not meet the \BI\ criterion.
However, the visually identified emergence continued to increase into the Gemini
observation, where it became strong enough (in all three quasars) to be
considered a BAL trough.


There were 11 cases in our sample of a quasar transitioning
from BAL to non-BAL, occurring in
10 different quasars (the quasar J015017 made this transition twice; see below).
Seven of these cases occurred in the BOSS-Gemini transition,
two in the SDSS1-SDSS2 transition, and two in the SDSS-BOSS transition.
The two cases of occurrence in the SDSS-BOSS transition were in J132508 and J150935,
and the spectrum of the former is shown in \S~\ref{sec:afteremerge}.

\begin{figure}[htb]
  \centering
  \includegraphics[width=0.75\columnwidth]{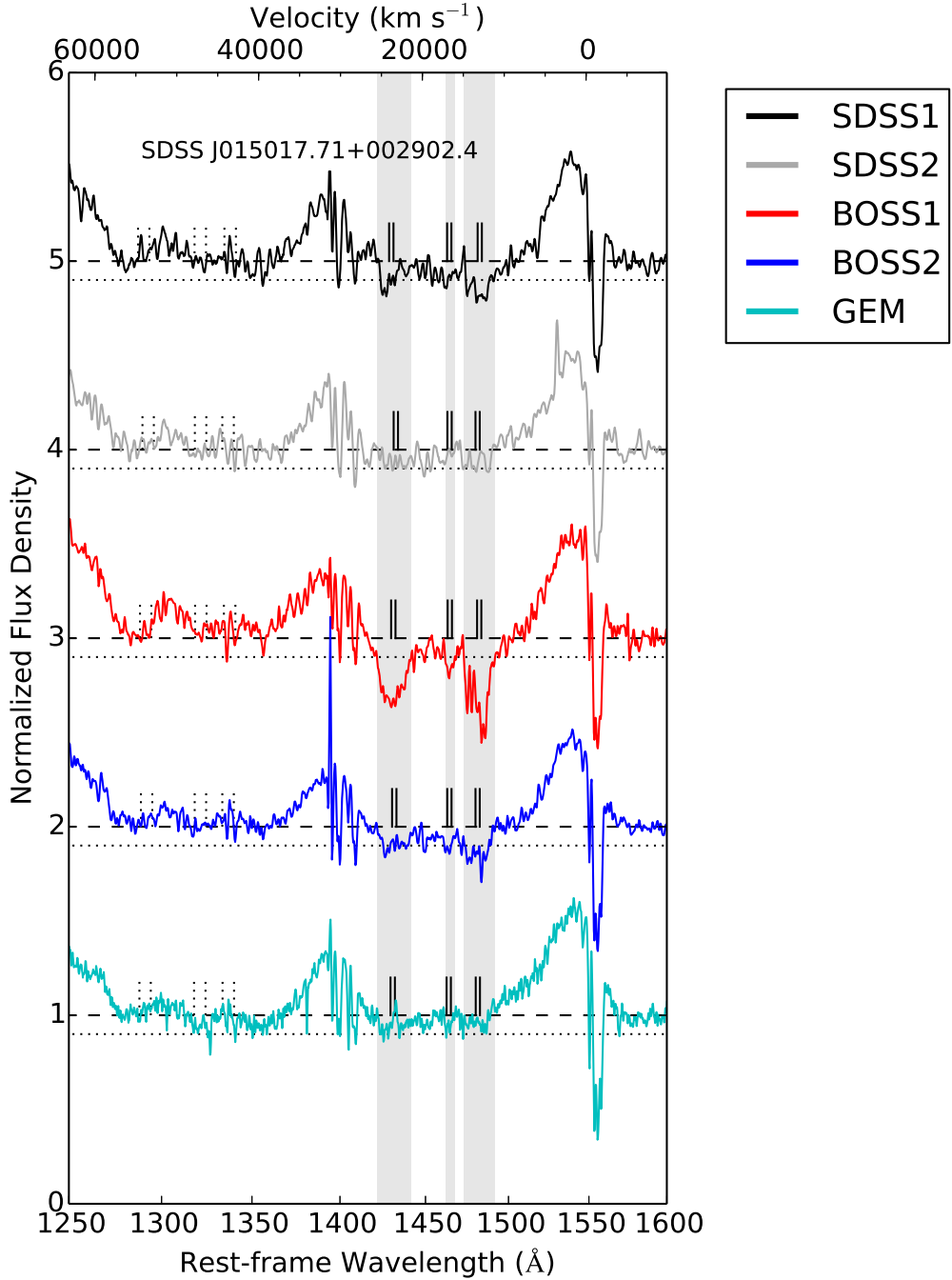}\hfill
  \caption{A plot of spectra taken at five different epochs for the quasar
  J015017. Note the changes in the three absorption complexes in
  the wavelength range 1400~\AA\ to 1500~\AA. With the first spectrum at the
  top, and each successive observation in chronological order below, we can
  observe these complexes emerge between the 2nd and 3rd observation, then
  disappear again between the 4th and 5th.}
  \label{fig:j015017}
\end{figure}

There were 5 quasars that exhibited both
emergence and disappearance over the course of our observations:
J015017, J022143, J081811, J095254, and J142054;
these objects have behavior similar to SDSS J093620.52+004649.2
\citep{2012ApJ...757..114F,2007ApJ...656...73L}.
Of particular interest was quasar J015017; its spectra are plotted in
Figure~\ref{fig:j015017}. In the 101.25 days between SDSS1 and SDSS2
observations, all BAL troughs disappeared in J015017. The troughs reappeared
with a much stronger absorption 745.54 days later in the BOSS1 observation.
Between BOSS1 and BOSS2 (66.41 days) the trough weakened substantially.
Finally, the troughs disappeared again in the Gemini observation, 303.05 days
after the last BOSS observation.


\subsection{BAL Quasar Emergence and Disappearance Rates} \label{sec:emergenceRate}

The candidate BAL trough emergence rate at $z>1.68$ for DR7$-$DR9 was
$111/7417=1.50\pm0.14$~\%, and for DR7$-$DR10 was
$181/8239=2.20\pm0.17$~\%.
These rates differ by 3.2$\sigma$ when only Poisson noise is considered
in the uncertainties.
Additional uncertainties in the rates can arise due to signal-to-noise-dependent identification of troughs, for example, so we consider the rates from the two subsamples to be consistent when all sources of uncertainty are considered.
Combined, we find a BAL trough emergence rate at $z>1.68$ of
$292/15656=1.87\pm0.11$~\%
over the rest-frame timescales of $1-3$~years separating DR7 and DR9+DR10.

The rate at which non-BAL quasars transitioned to BAL quasars is lower,
as many BAL troughs appeared in pre-existing BAL quasars.
From the previous section, 33 non-BAL to BAL transitions occurred between
SDSS and BOSS among 103 visually identified candidates followed up with Gemini spectra, a fraction of $32 \pm 6$~\%.
Thus, we find a non-BAL to BAL quasar emergence rate of $R_E = 0.59 \pm 0.12$~\%
at $z>1.68$ over rest-frame timescales of $1-3$~years.

In equilibrium, the BAL emergence and disappearance rates $R_E$ and $R_D$
and the numbers of BAL and non-BAL quasars $N_B$ and $N_N$ in the parent sample
are related by $R_E N_N = R_D N_B$, or $R_E = R_D (N_B/N_N)$.

In \citet{2012ApJ...757..114F}, 21 \civ\ broad absorption features were observed
to disappear in 19 quasars selected from a parent sample of 582 BAL
quasars in SDSS and BOSS.
In 10 of those cases the quasars transformed from BAL to non-BAL,
for a rate of $R_D=1.7^{+0.7}_{-0.5}$~\% of BAL quasars disappearing
(transforming into non-BAL quasars) over SDSS-BOSS timescales.

Using the above values for $R_D$ and $R_E$, we find
$R_E/R_D = N_B/N_N = 0.35^{+0.12}_{-0.16}$.
That BAL to non-BAL quasar ratio $r\equiv N_B/N_N$ corresponds to a BAL quasar
fraction in the parent sample of $f_{BAL}=r/(1+r)=0.26^{+0.09}_{-0.12}$.
That value is consistent within the uncertainties
with the incompleteness-corrected BAL quasar fraction
of $f_{BAL} = 0.140\pm 0.016$ found by \citet{allenbal}
at $z\geq 1.5$ in the SDSS.
However, their smaller observed value of $f_{BAL}$ predicts
a smaller emergence/disappearance ratio of
$r=R_E/R_D=f_{BAL}/(1-f_{BAL})=0.16\pm 0.02$; in other words, either a lower emergence
rate than we observe or a higher disappearance rate than observed in
\citet{2012ApJ...757..114F}, or both.
Nonetheless, our BAL quasar emergence rates
and the BAL quasar disappearance rates of \citet{2012ApJ...757..114F}
are in agreement within the uncertainties.

\subsection{Changes in Equivalent Width}\label{sec:deltaEW}

With 103 quasars with 219 absorption complexes, and at least 3 observations
per quasar, there are 526 epoch-to-epoch changes in EW, depth, and centroid
velocity. In that set of data, 462 of the changes were statistically
significant at greater than 3$\sigma_{\textrm{EW}}$, while the other 64 were
within the noise. Thus, the fraction of absorption complexes that exhibited
a statistically significant change in EW was 462/526=88$\pm$4~\%, and no
change was 64/526=12$\pm$1~\%. Of course, this is biased
because we specifically chose quasars to observe that showed some obvious
changes in their spectra between the first two observations.

\begin{figure}[htb]
    \centering
    \includegraphics[scale=0.575]{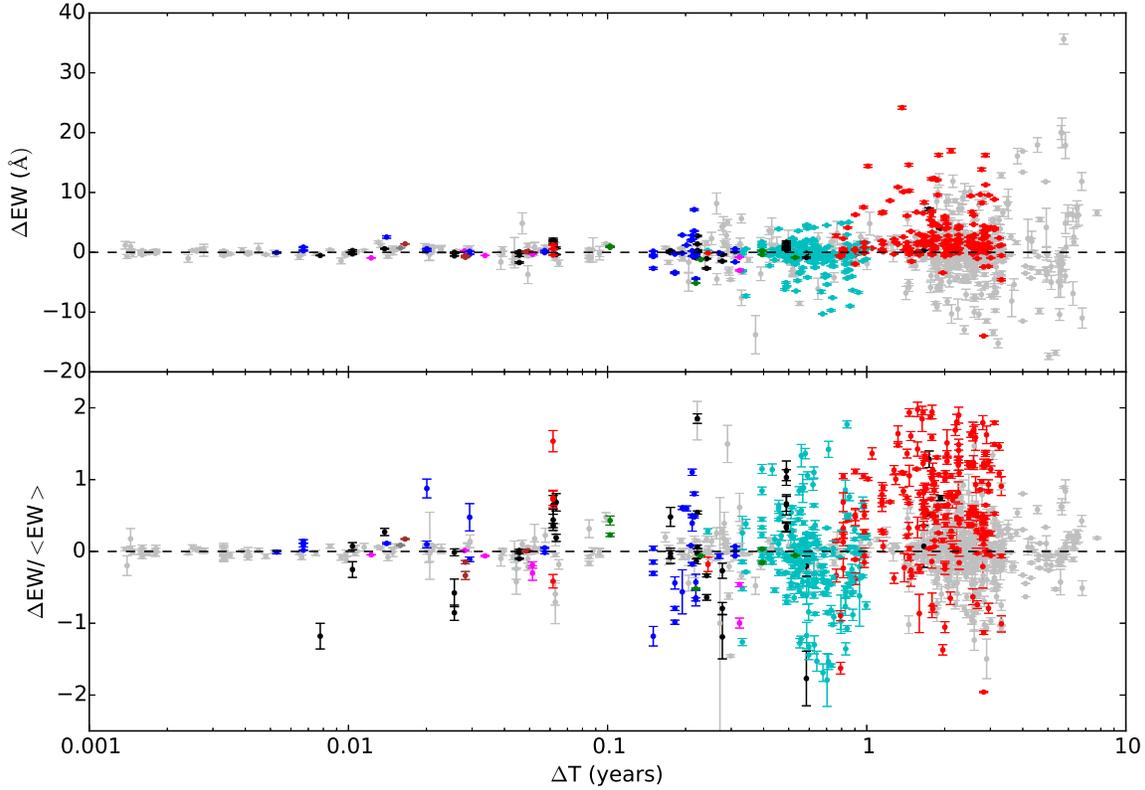}
    \caption{The change in equivalent width, $\Delta$EW, (top) and change in fractional
    equivalent width, $\Delta \textrm{EW} / \langle \textrm{EW} \rangle$,
    between successive epochs in a quasar as a function of rest-frame time
    between them.
    The grey points are taken from the literature (see Table~\ref{tab:VarLit}).
    All other points are from this work. Black points represent changes between
    two SDSS observations, red points are changes from SDSS to BOSS
    observations, blue are changes between two BOSS observations, cyan are
    changes from BOSS to Gemini observations, and all other colors are
    changes between two Gemini observations.}
    \label{fig:deltaEW}
  \end{figure}

We plot the change in EW between successive epochs of all
absorption complexes in the top portion of Figure~\ref{fig:deltaEW} (see
equation \ref{eq:deltaEW}), and the fractional change in the bottom portion
(see equation \ref{eq:frac}).
All of the grey points in both figures are values taken from the literature
(see Table~\ref{tab:VarLit}). The colored points are from this work:
black points represent changes between two SDSS observations, red points are
changes from SDSS to BOSS
observations, blue are changes between two BOSS observations, cyan are changes
from BOSS to Gemini observations,
and all other colors are changes between two Gemini observations.
We note that our data largely traces the results of other works: on longer
time-scales, large changes in EW are possible. There are no large changes in
EW on short time-scales. Figure~\ref{fig:deltaEW} also shows how our dataset
has filled in the region on medium time-scales of $0.5-1.0$ years in the
rest-frame, a region of parameter space that has been largely unexplored.
Both absolute and relative EW change measurements are useful.  The quantity $\Delta$EW/$\langle$EW$\rangle$ directly identifies BAL emergence ($\Delta$EW/$\langle$EW$\rangle$ = 2) and disappearance ($\Delta$EW/$\langle$EW$\rangle$ = $-2$).  BAL emergence can occur on timescales as short as 0.2 rest-frame years, but only relatively small $\Delta$EW values are observed on such timescales.
On timescales longer than 1 rest-frame year, our SDSS-BOSS comparisons (red points) show a larger range of both absolute and relative EW changes but are not dominated by cases of pure emergent troughs.  Instead, strengthening troughs with the full range of positive $\Delta$EW/$\langle$EW$\rangle$ values are observed.

\begin{figure}[htb]
  \centering
  \includegraphics[scale=0.475]{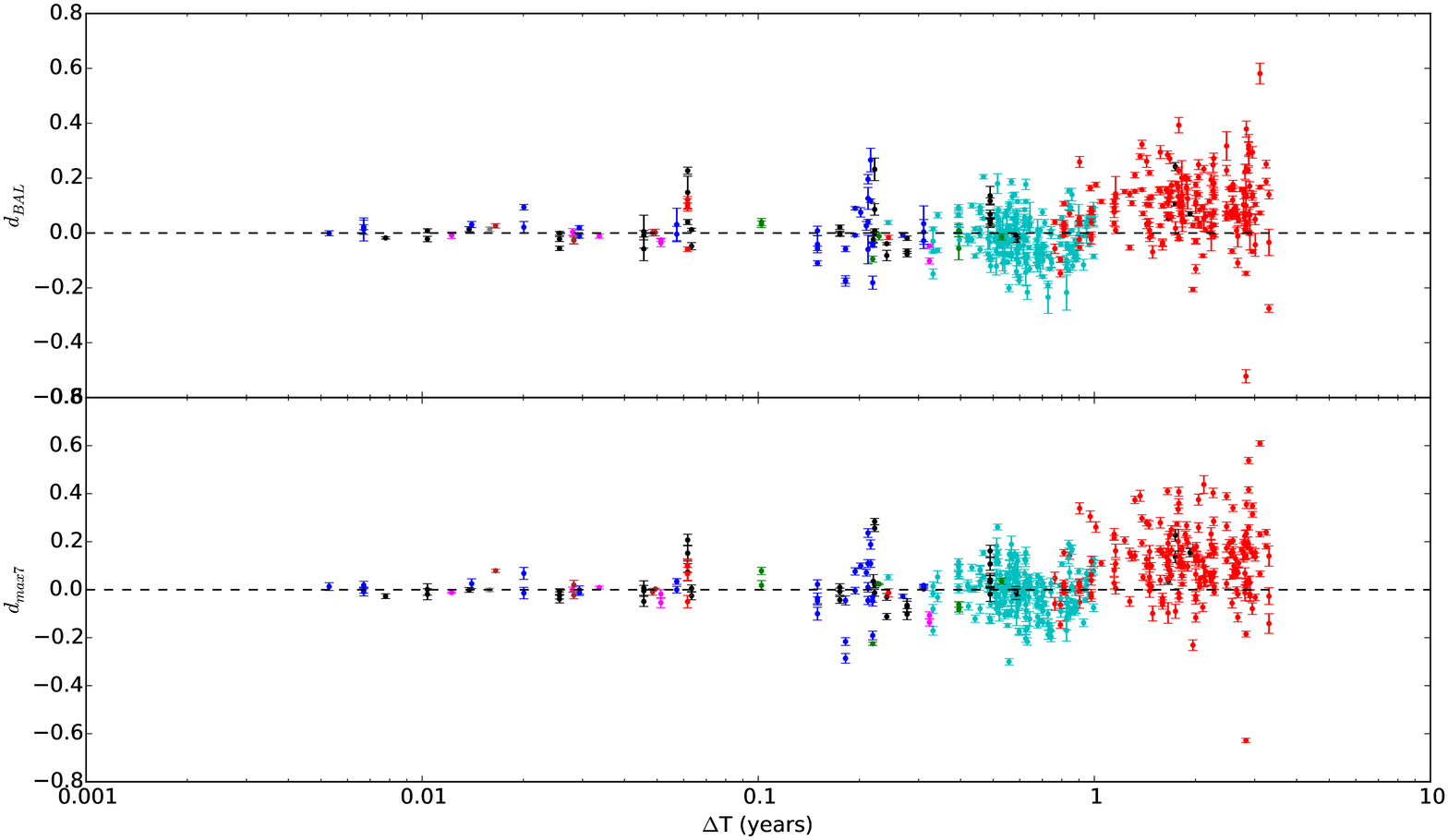}
  \caption{The change in $d_{BAL}$ (top) and $d_{max7}$ (bottom) between successive
  epochs in a quasar as a function of rest-frame time between observations.
  Black points represent changes between
  two SDSS observations, red points are changes from SDSS to BOSS
  observations, blue are changes between two BOSS observations, cyan are
  changes from BOSS to Gemini observations, and all other colors are
  changes between two Gemini observations.}
  \label{fig:deltaD}
\end{figure}

In Figure~\ref{fig:deltaD}, the change in depth of trough is plotted
against the time between successive observations as measured by $d_{BAL}$
(top) and by $d_{max7}$. Both figures display a similar trend of
EW changes with time: on longer time-scales, larger changes are possible.
The similarity between Figure~\ref{fig:deltaD} and Figure~\ref{fig:deltaEW} is due to the fact that most EW variability is dominated by changes in trough depth rather than by changes in trough width.

\subsection{Comparison of BAL quasars with and without emergent troughs} \label{sec:DR12QBAL}


To study the properties of pre-existing BAL quasars and emergent BAL quasars in our DR7-DR9 and DR7-DR10 samples, we matched our samples with the SDSS DR12 BAL quasar catalog \citep{2017A&A...597A..79P}, hereafter referred to as DR12QBAL.
The DR12QBAL catalog consists of automated detections and measurements of BAL troughs in each DR12 quasar visually identified as having BAL features in its highest-SNR BOSS spectrum.

The DR12QBAL catalog measurements include the traditional balnicity index, here denoted $\rm BI_{\rm 2000}$:
\begin{equation}
{\rm BI_{\rm 2000}} = \int_{3000}^{25000}\left ( 1 - \frac{f(v)}{0.9} \right ) C_{2000} ~dv,
\label{eq:BI2000}
\end{equation}
where $C_{2000}=1$ when the quantity in parentheses for a candidate trough is positive for more than 2000 km s$^{-1}$.
The DR12QBAL catalog also includes the absorption index \citep{sdss123}:
\begin{equation}
{\rm AI_{\rm 450}} = \int_{0}^{25000}\left ( 1 - \frac{f(v)}{0.9} \right ) C_{450} ~dv,
\label{eq:AI450}
\end{equation}
where $C_{450}=1$ when the quantity in parentheses for a candidate trough is positive for more than 450 km s$^{-1}$.

For each visually identified BAL quasar, the DR12QBAL catalog contains total $\rm BI_{\rm 2000}$ and $\rm AI_{\rm 450}$ values and the number of $\rm BI_{\rm 2000}>0$ and $\rm AI_{\rm 450}>0$ troughs; for each trough, velocity limits, maximum depths, and maximum depth velocities are given.
DR12QBAL does not contain the $\rm BI_{\rm 2000}$ and $\rm AI_{\rm 450}$ values of the individual troughs in each object.
We estimated those values as follows.
We estimated the relative EW of each trough as the product of its velocity width times its maximum depth, divided by the sum of those products for all troughs in each object.  While the individual EWs are clearly overestimates, the relative EWs are our best estimates of the relative strength of each trough.  We multiply each trough's relative EW by the total $\rm BI_{\rm 2000}$ or $\rm AI_{\rm 450}$ of the quasar to each trough's estimated $\rm BI_{\rm 2000}$ or $\rm AI_{\rm 450}$.

From our DR7-DR9 sample of 8317 quasars at $z>1.68$, we find
1162 DR7-DR9 quasars in DR12QBAL for which measurements are available on the same BOSS spectra, including 77 of our emergent BAL quasar candidates.


From our DR7-DR10 sample of 8239 quasars at $z>1.68$, we find
1334 DR7-DR10 quasars in DR12QBAL for which measurements are available on the same BOSS spectra, including 99 of our emergent BAL quasar candidates.

We chose not to compare our DR9-DR10 quasar sample with DR12QBAL, because of the significantly shorter timescales between the first two epochs in that sample.
This exclusion only affects five quasars with Gemini spectra.

We therefore have information from DR12QBAL
on 176 emergent BAL quasar candidates,
including 51 with Gemini spectra,
and 2320 non-emergent BAL quasars.

There are 116 emergent BAL quasar candidates with absorption undetected in DR12QBAL, including 49 with Gemini spectra.
Approximately half of those 49 non-detections are due to the emergent BAL trough being at a velocity $>$ 25000 \kms.  The rest are a roughly equal mixture of shallow and narrow troughs.  The differences between the catalogs in their common velocity range could be due to differences in continuum fitting, smoothing before BAL detection, and visual inspection of BOSS spectra alone for DR12QBAL as opposed to visual inspection of DR7 and BOSS spectra simultaneously for our catalog.

%

Next, in the 51 Gemini targets also found in the DR12QBAL catalog, we attempted to match our trough complexes with DR12QBAL troughs with $\rm BI_{\rm 2000}>0$.  We chose to study matches with $\rm BI_{\rm 2000}>0$ troughs because our 1000 \kms\ minimum trough width means we should detect all such troughs, but not all troughs with $\rm AI_{\rm 450}$.
In this matching, we identified all $\rm BI_{\rm 2000}>0$ troughs overlapping in velocity with one of our trough complexes as part of that trough complex.  We calculated the $\rm BI_{\rm 2000}$ value, velocity width, maximum depth, and velocity of maximum depth for each trough complex.

Only one trough with $\rm BI_{\rm 2000}>0$ in the 51 Gemini targets also found in the DR12QBAL catalog did not have a counterpart among our trough complexes.  It is a trough in J123404 on the shoulder of the \civ\ broad emission line.  It did not enter our catalog because we did not include the broad emission lines in our normalizing continuum estimate.



There were 23 trough complexes detected by us in those 51 Gemini targets which were not detected in the DR12QBAL catalog (even with $\rm AI_{\rm 450}>0$).  They are a roughly equal mixture of complexes outside the DR12QBAL velocity range, complexes where a reasonable difference in continuum placement could remove the trough detection, and complexes for which there is no obvious explanation for the oversight in DR12QBAL without knowing the continuum used by DR12QBAL.

\begin{figure}[htb]
  \centering
  \includegraphics[width=0.725\columnwidth]{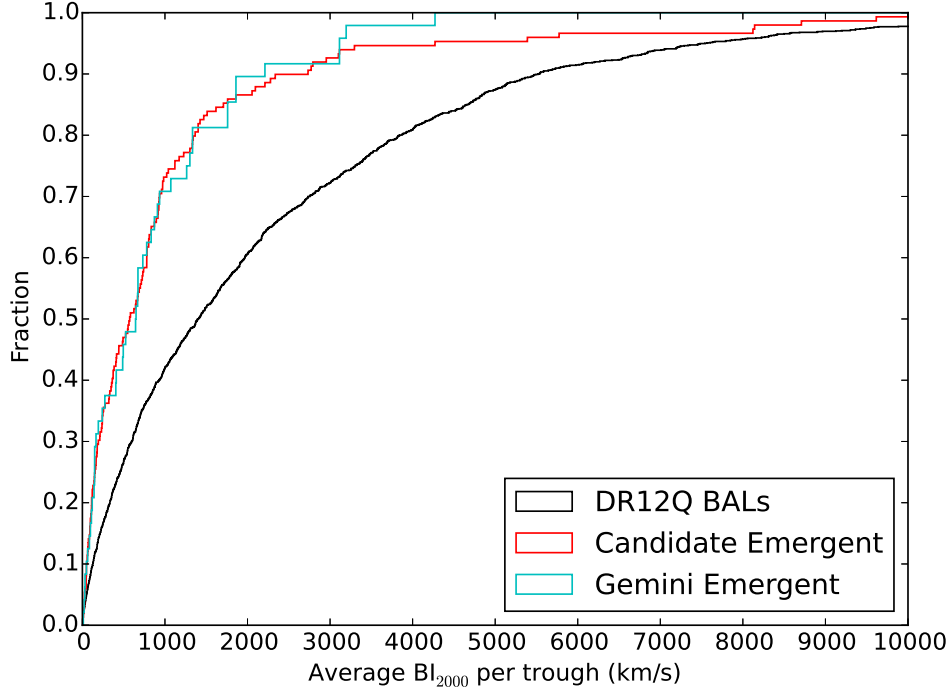}
  \caption{The cumulative distribution functions (CDFs) of average BI$_{\rm 2000}$ values per trough in non-emergent DR12Q BAL quasars (black), candidate emergent DR12Q BAL quasars (red), and emergent DR12Q BAL quasars with Gemini spectra (cyan).}
  \label{fig:cdfBI2000}
\end{figure}
\begin{figure}[htb]
  \centering
  \includegraphics[width=0.725\columnwidth]{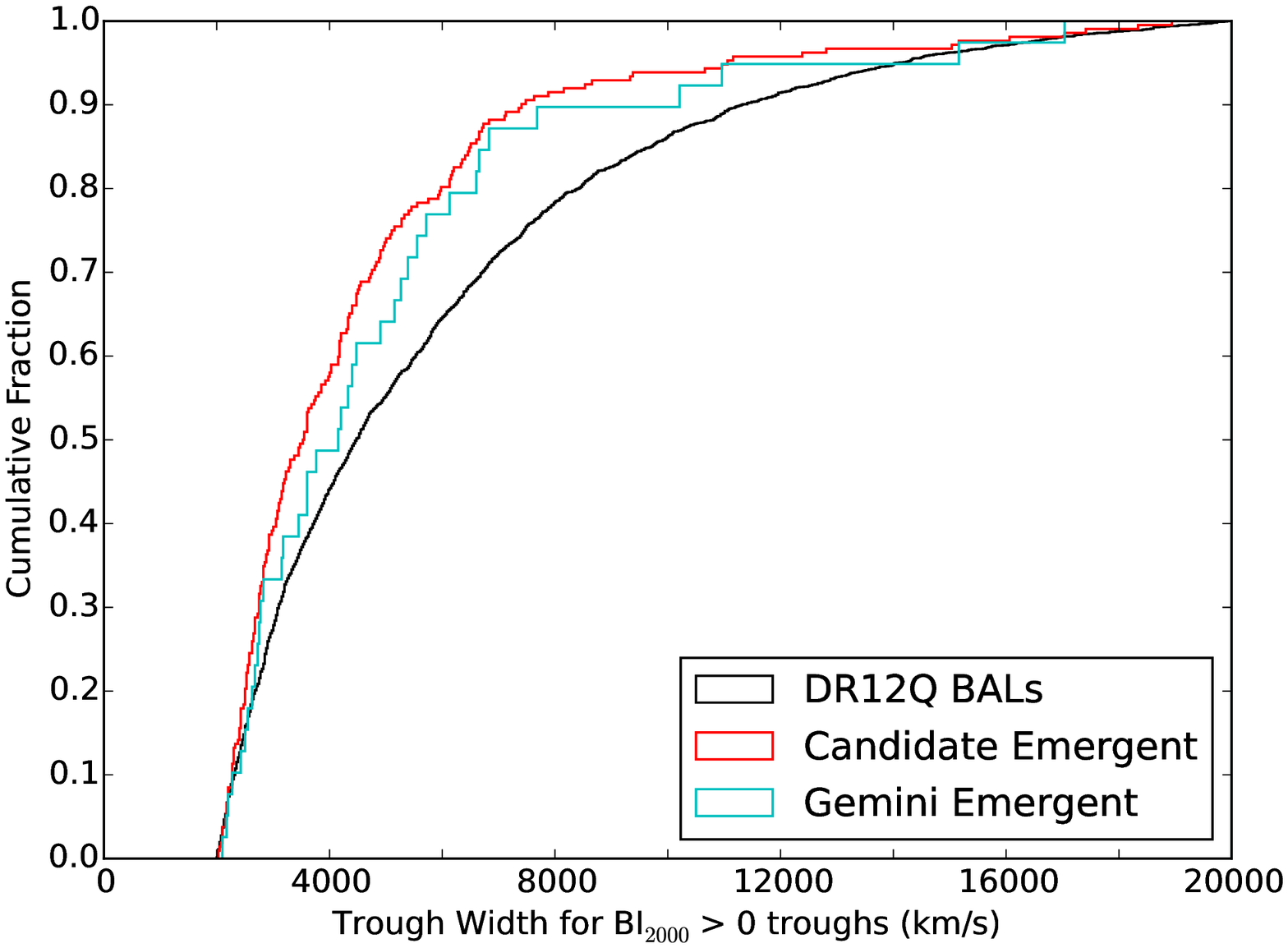}
  \caption{The cumulative distribution functions (CDFs) of average DR12Q BAL trough widths in non-emergent DR12Q BAL quasars (black), candidate emergent DR12Q BAL quasars (red), and emergent DR12Q BAL quasars with Gemini spectra (cyan).}
  \label{fig:cdfBIwidth}
\end{figure}
\begin{figure}[htb]
  \centering
  \includegraphics[width=0.725\columnwidth]{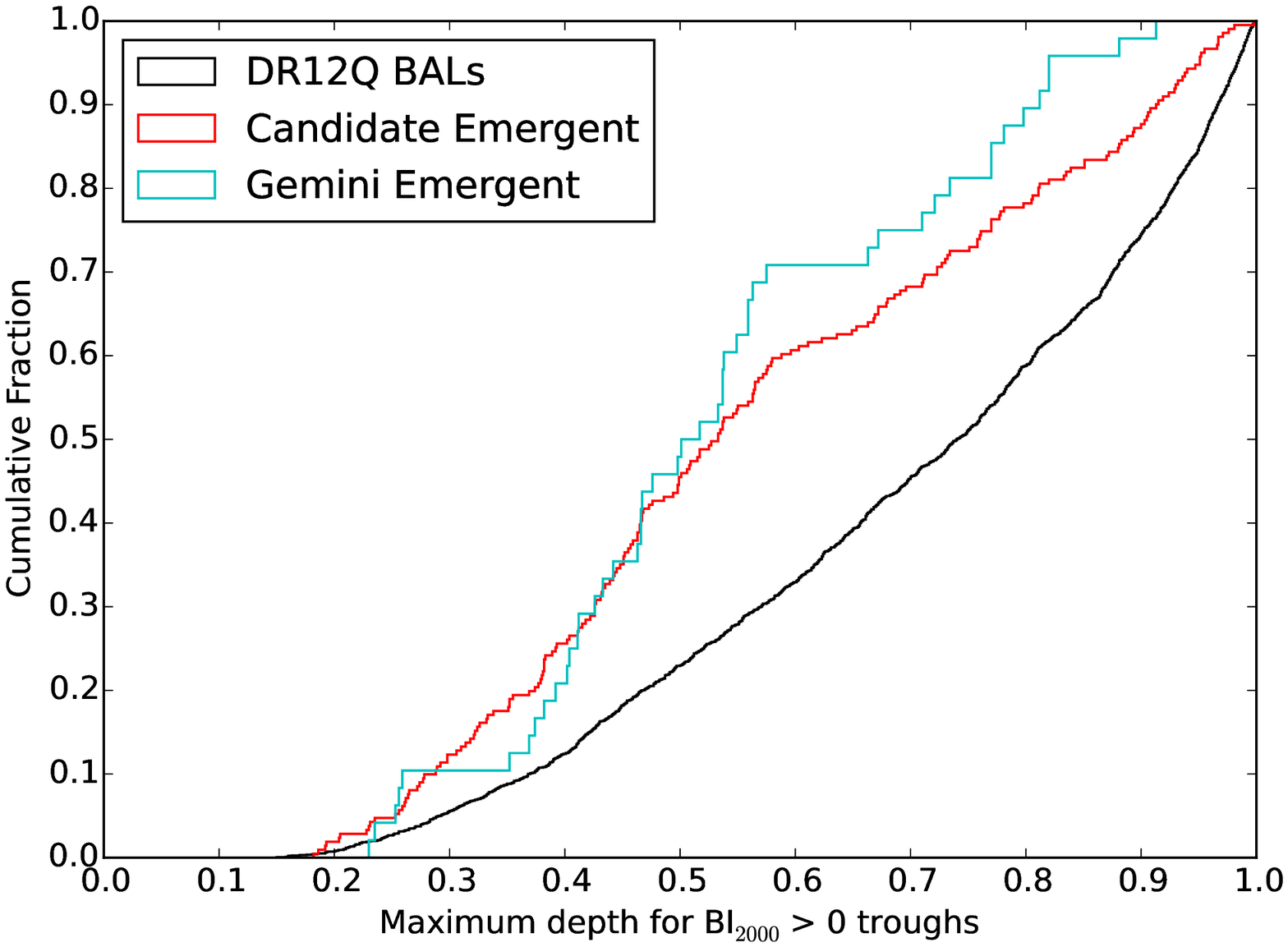}
  \caption{The cumulative distribution functions (CDFs) of DR12Q BAL trough depths in non-emergent DR12Q BAL quasars (black), candidate emergent DR12Q BAL quasars (red), and emergent DR12Q BAL quasars with Gemini spectra (cyan).}
  \label{fig:cdfBIdepth}
\end{figure}
\begin{figure}[htb]
  \centering
  \includegraphics[width=0.725\columnwidth]{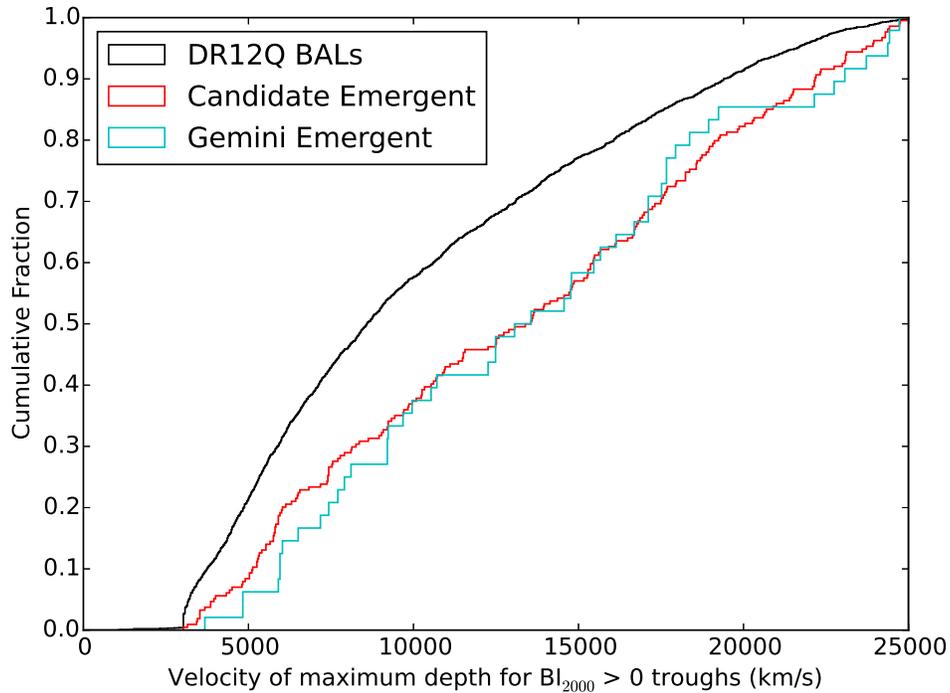}
  \caption{The cumulative distribution functions (CDFs) of the velocities of DR12Q BAL trough maximum depths in non-emergent DR12Q BAL quasars (black), candidate emergent DR12Q BAL quasars (red), and emergent DR12Q BAL quasars with Gemini spectra (cyan).}
  \label{fig:cdfBIvmaxdepth}
\end{figure}

In Figures \ref{fig:cdfBI2000} to \ref{fig:cdfBIvmaxdepth}, we show the cumulative distribution functions (CDFs) for trough $\rm BI_{\rm 2000}$, velocity width, maximum depth, and velocity of maximum depth for emergent and non-emergent BAL quasar samples with information from DR12QBAL.
In these figures it can be seen that the histograms from the 51-object Gemini subsample (cyan) are close to those from the 176-object candidate emergent subsample (red).
Those histograms both differ from those of the 2320-object non-emergent DR12QBAL quasar sample (black).

To compare these CDFs quantitatively, we use the
two-sample Kuiper variant of the Kolmogorov-Smirnov (K-S) test
\citep{nr3}.\footnote{The sensitivity of the traditional K-S test
diminishes at the extreme ends of the CDFs (i.e., at $P(x)$ = 0 or 1,
where $x$ is the independent variable). The Kuiper statistic eliminates
the diminished sensitivity by summing the maximum values of the differences
in two CDFs in both directions. Thus, the Kuiper statistic V is given by
$V = |\max(P_2(x)-P_1(x))|+|\max{P_1(x)-P_2(x)}|$.}
The results are provided in Table~\ref{tab:KStestSamples}.
In each parameter, the Candidate Emergent sample is statistically distinguishable from the DR12QBAL sample at high significance.
For the first two parameters in the Table, the Gemini Emergent subsample is
statistically representative of the Candidate Emergent sample;
we cannot make that claim for the other two parameters.

Overall, we find that candidate newly emerged BAL quasar troughs are preferentially drawn from among BAL troughs with smaller BI$_{\rm 2000}$ values, shallower depths, larger velocities, and smaller widths.

\begin{table}
    \caption{The p$-$values calculated using the Kuiper variant of the Kolmogorov–Smirnov test for each BAL parameter represented in Figures~\ref{fig:cdfBI2000} through \ref{fig:cdfBIvmaxdepth}. The percentages indicate the probability that two given distributions are drawn from the same parent sample. }
    \label{tab:KStestSamples}
    \begin{tabular}{lll}
      \hline \hline
      BAL parameter & Candidate Emergent vs. DR12Q (\%) & Gemini Emergent vs. Candidate (\%) \\
      \hline
      BI$_{\rm 2000}$ & $1.268\times10^{-9}$ & 99.99 \\
      Trough width & $1.925\times10^{-3}$ & 95.26 \\
      Trough depth & $8.776\times10^{-11}$ & 26.25 \\
      Velocity of maximum depth & $7.331\times10^{-5}$ & 69.65 \\
      \hline
    \end{tabular}
\end{table}

\clearpage

\subsection{What Happens to the Absorption After it Emerges?}\label{sec:afteremerge}

\begin{figure}[htb]
  \centering
  \includegraphics*[scale=0.75]{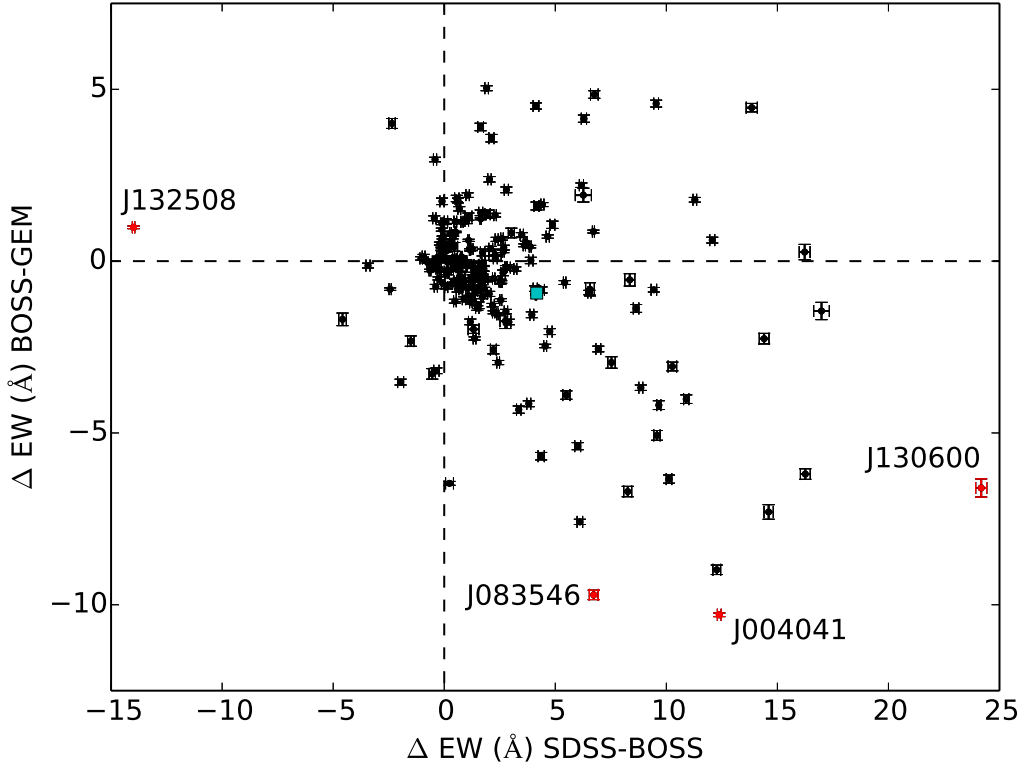}
  \caption{Comparing the change in EW between the SDSS and BOSS observations to
  the same quasar's change in EW between the BOSS and GEM observations.
  The 4 red points are the extreme cases shown in Fig.~\ref{fig:weird}.
  The cyan square is the weighted-mean value of all the points in the plot.}
  \label{fig:sbbg}
\end{figure}

Our original candidate sample was targeted for visual identification of
emergent absorption. Thus our sample is biased toward an increase in
EW when comparing an SDSS observation to a BOSS observation.
In Figure~\ref{fig:sbbg}, the change in EW ($\Delta$EW) measured between SDSS
and BOSS observations is
compared to the change in the same absorption complex between the BOSS
observation and our Gemini observation. Note that there are some quasars with
multiple SDSS and/or BOSS observations, but the figure only plots the two
with the shortest separation in time (i.e., SDSS2-BOSS1); in target selection,
these were the spectra that were visually compared. Thus, for each
absorption complex, we have one point for the figure. There are 3 exceptions:
the quasars that were not discovered until SDSS Data Release 9 (see \S~\ref{sec:obsTable}); they are not included in the plot.

\begin{figure}[htb]
\centering
  \begin{tabular}{@{}cc@{}}
    \includegraphics[width=.34\textwidth]{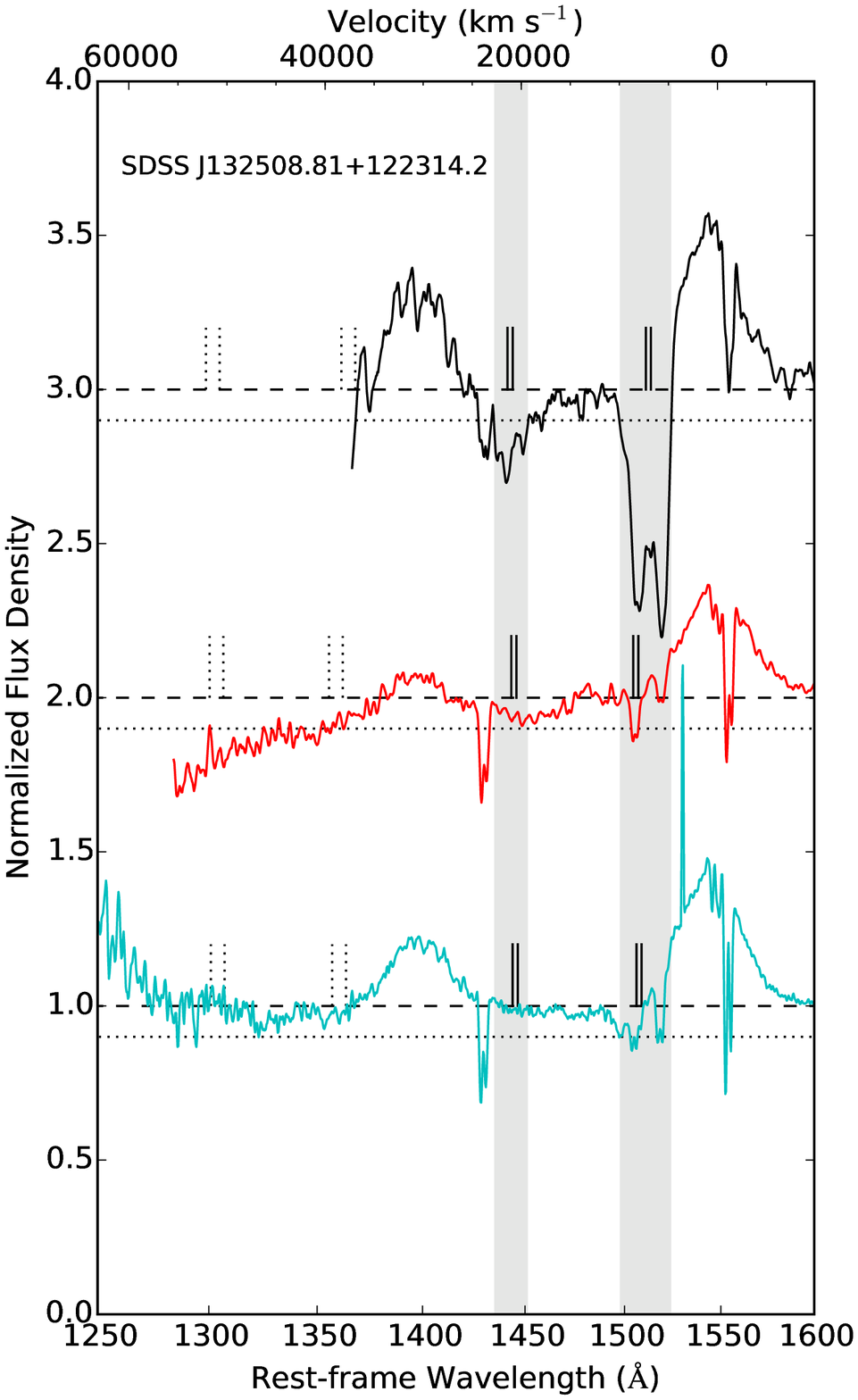} &
    \includegraphics[width=.33\textwidth]{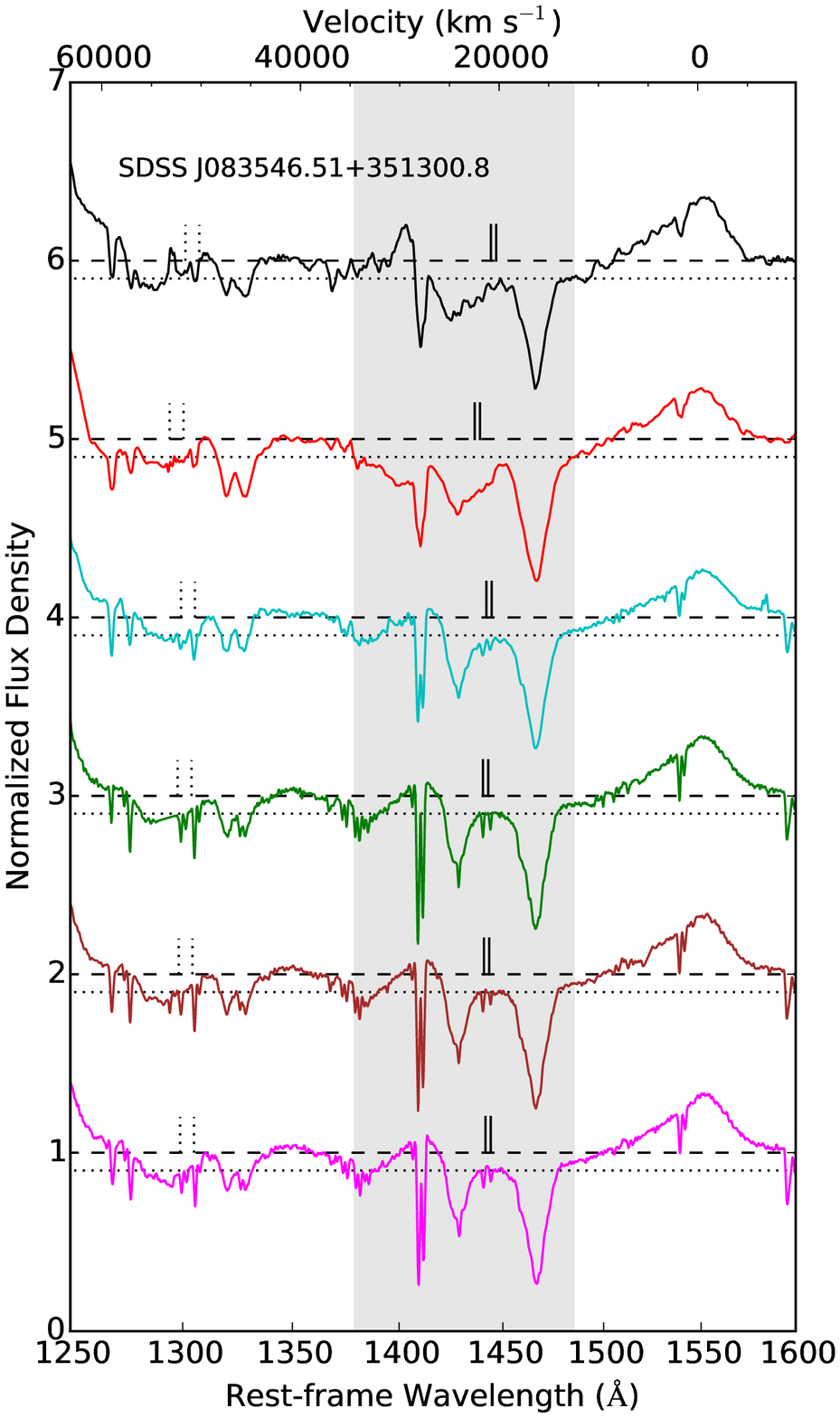} \\
    \includegraphics[width=.34\textwidth]{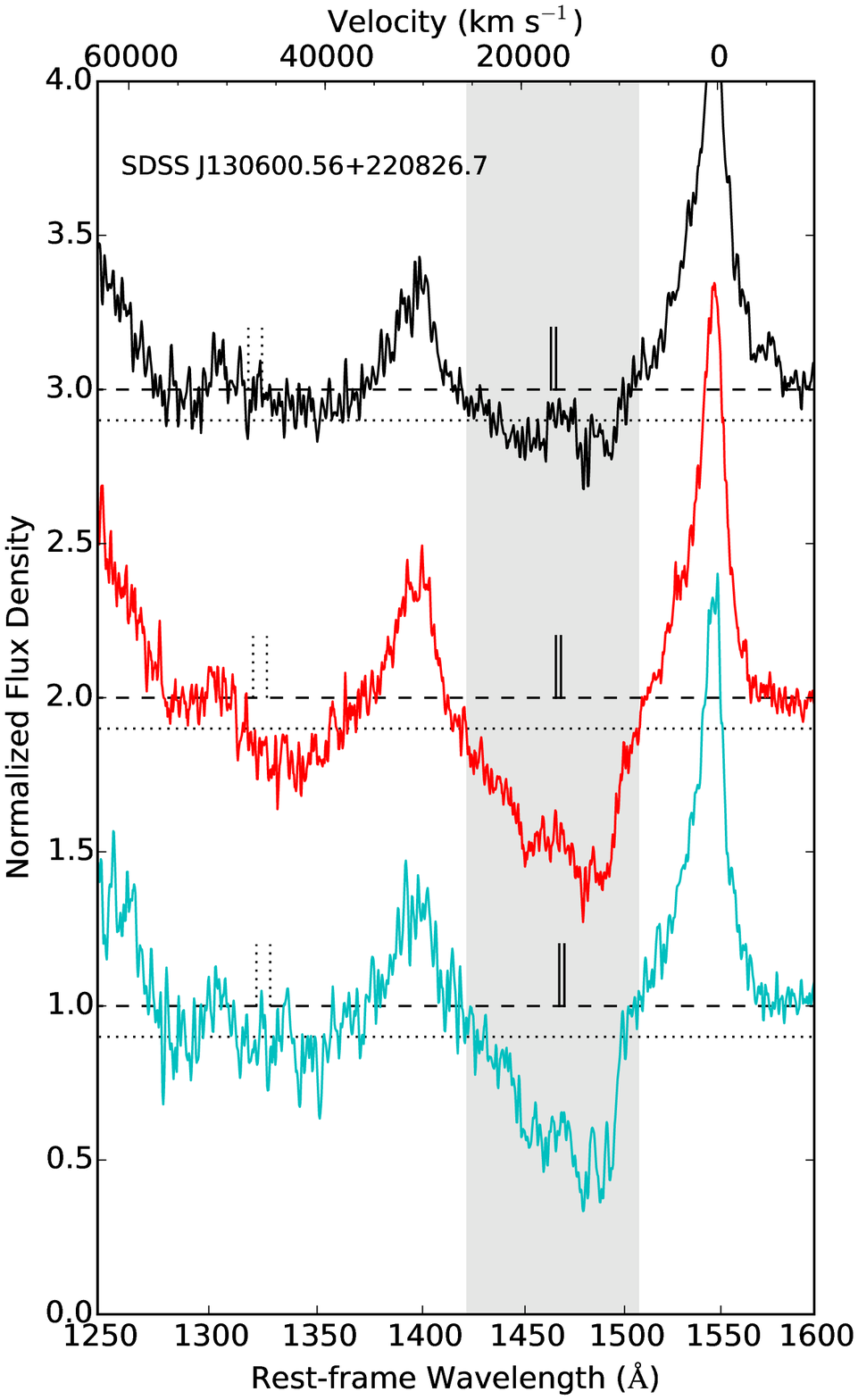} &
    \includegraphics[width=.33\textwidth]{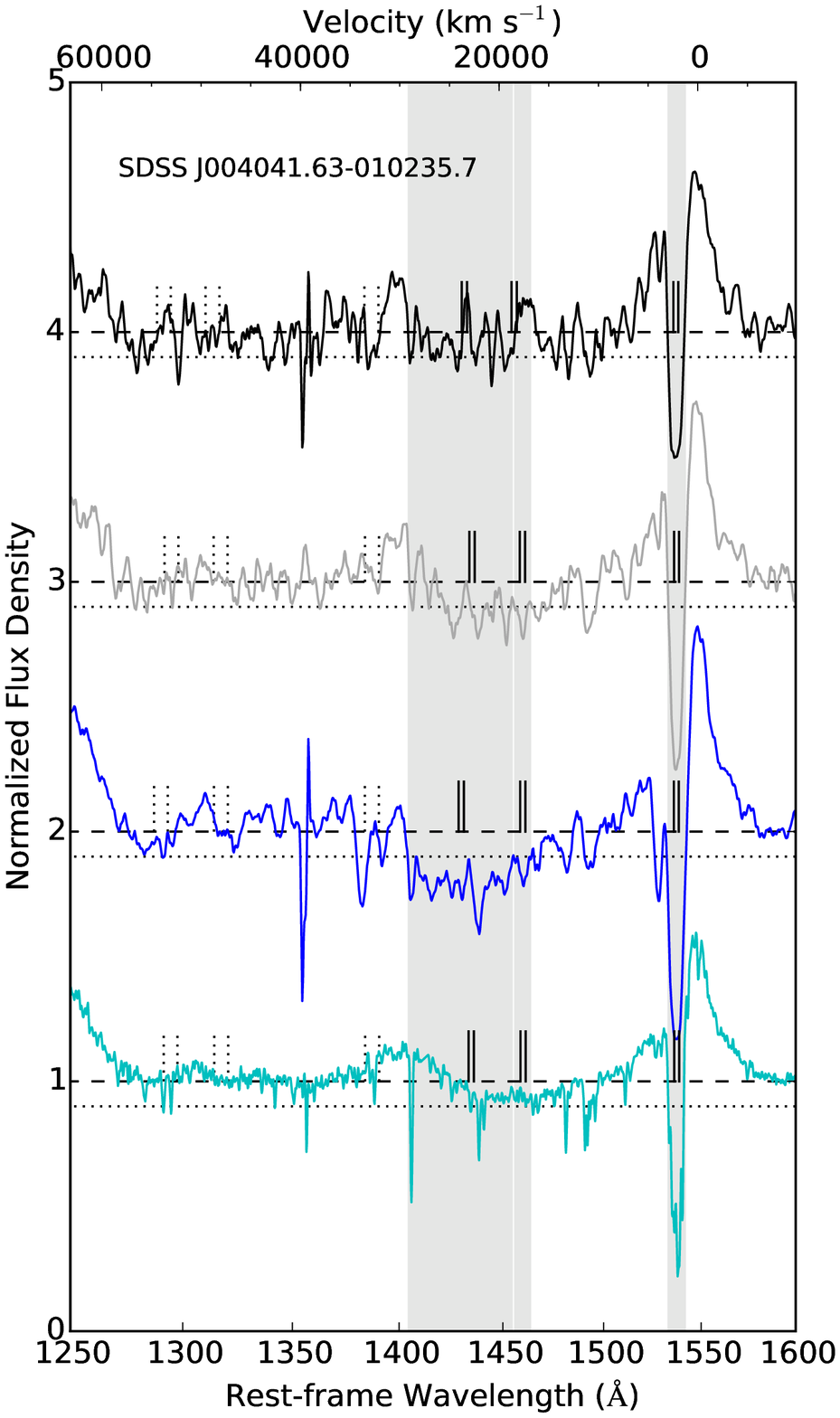}
  \end{tabular}
  \caption{The four quasars highlighted as red points in Fig.~\ref{fig:sbbg}
  represent some of the most extreme changes in absorption measured in this
  dataset. Each quasar's spectra are plotted, with the oldest spectrum is at
  the top with each successive spectrum in chronological order below it.
  Colors represent the differing observational origins of each spectrum.
  See Appendix~\ref{app:figures} for more information on these figures.}
  \label{fig:weird}
\end{figure}

Most of the data points are found to the right of
$\Delta \textrm{EW SDSS-BOSS} =0$, indicating an absorption
feature's EW usually but not always increased between SDSS and BOSS observations.
This result is expected because in building the visually identified
emergent sample in \S~\ref{sec:targetSelection}, we only searched for new
troughs, not new BAL quasars; thus, we expect a small number of weakening
pre-existing troughs in a sample with most objects having increasing EW.
Further, $\sim$~60\%\ of the data points are below $\Delta \textrm{EW BOSS-GEM} =0$, indicating the EW of an absorption feature usually got smaller
between the BOSS and Gemini observations.
The cyan point shows the weighted-mean values of $\Delta$EW of $4.16\pm0.10$~\AA\ from SDSS to BOSS and $-0.93\pm0.09$~\AA\ from BOSS to Gemini.
The four red points in Figure~\ref{fig:sbbg} are highlighted because they exhibit extreme changes in EW between both SDSS to BOSS observations and BOSS to Gemini
observations. Their spectra are plotted in Figure~\ref{fig:weird}. J132508
exhibited a large decrease in EW from SDSS to BOSS
observations. The spectra indicate that a strong absorber
at small velocities completely disappeared between the two observations.
J132508 was originally targeted for a possible absorber
emerging at $v > 40,000$~\kms, but the larger wavelength coverage
of the Gemini observation showed that candidate absorber to be spurious.
J130600 exhibits some of the strongest absorption seen in our entire dataset,
and the strongest increase in EW from SDSS to BOSS.
In the BOSS spectrum of J083546 there is strengthening of the \civ\ absorption on top of its \siv\ emission, but the absorption was weaker in the first Gemini epoch (cyan) than in the original SDSS epoch.
Finally, J004041 exhibited a large, albeit shallow, increase in \civ\
absorption between SDSS2 (grey) to BOSS (blue); this increase was almost
completed reversed in the Gemini epoch (cyan).

\clearpage

\subsubsection{Conditional probabilities of absorption changes} \label{sec:conditional}

We now consider in more detail how a previous change in equivalent
width could predict the change in EW between the next two epochs.
We are interested in how troughs
behave after a change has been observed. If a trough changes
its EW, what is it most likely to do next? Will it continue to change in
the same way? Or will it reverse its change? And on what time-scales?
Our goal is to determine how well a history of increases
(or decreases) in absorption could predict what the absorption would do next.

We utilize our full suite of 3-7 epochs for each quasar. For a given quasar,
we first compare the EWs between the first and second observations.
The EW change can either be an increase, a decrease, 
or within the uncertainties (i.e., no statistically significant change).
Next we compare the change in EW between the second and third observations and
determine if it was an increase, a decrease, or within the uncertainties.
If both changes in EW were increases or both changes in EW were decreases,
we label the EW trend `staying the same'.
If the first change in EW is an increase but the second is a decrease (or vice versa), we label the EW trend `changed'.
If the EW changes are within the uncertainties we label the EW trend `uncertain'.

There were, however, some caveats to the above prescription.
Below is an itemized list of how we built
this analysis, assuming a quasar has been observed $n$ times
(i.e., epoch$_1$\nodata epoch$_n$).

\begin{enumerate}
  \item{Determine the reference point.
  The reference point is the first change of EW from one epoch to another
  that is significant at $\geq 3\sigma_{EW}$.
  Start with the first observation, epoch$_1$, and search for a reference
  point in this order: epoch$_1$-epoch$_2$, epoch$_2$-epoch$_3$,
  epoch$_1$-epoch$_3$, epoch$_3$-epoch$_4$.}
  \item{If epoch$_1$-epoch$_2$ is SDSS1-SDSS2, ignore it unless
  epoch$_2$-epoch$_3$ showed no statistically significant change in EW. As
  our visual sample was built around searching for emergence between SDSS-BOSS,
  we set that to be the reference point as much as possible.}
  \item{Determine the sign of the change in EW from the reference point, either
  positive for increasing absorption or negative for decreasing absorption.}
  \item{Compare the reference point to the next epoch change. For example,
  if the reference point was epoch$_n$-epoch$_{n+1}$, determine whether the
  change in EW over epoch$_{n+1}$-epoch$_{n+2}$ is in the same direction or
  opposite to the reference point. Record the $\Delta T$ between
  epoch$_{n+1}$-epoch$_{n+2}$. However, if no statistically significant change
  in EW occurred between epoch$_{n+1}$-epoch$_{n+2}$, then compare the
  reference point to epoch$_{n+1}$-epoch$_{n+3}$, epoch$_{n+1}$-epoch$_{n+m}$,
  etc., until a significant change is found.}
  \item{Reset the reference point to epoch$_{n+1}$-epoch$_{n+m}$, the first two epochs between which a statistically significant change in EW was found to occur.}
  \item{Return to step 3 and repeat until there are no more epochs.}
\end{enumerate}

\begin{figure}[htb]
  \centering
  \includegraphics*[scale=0.80]{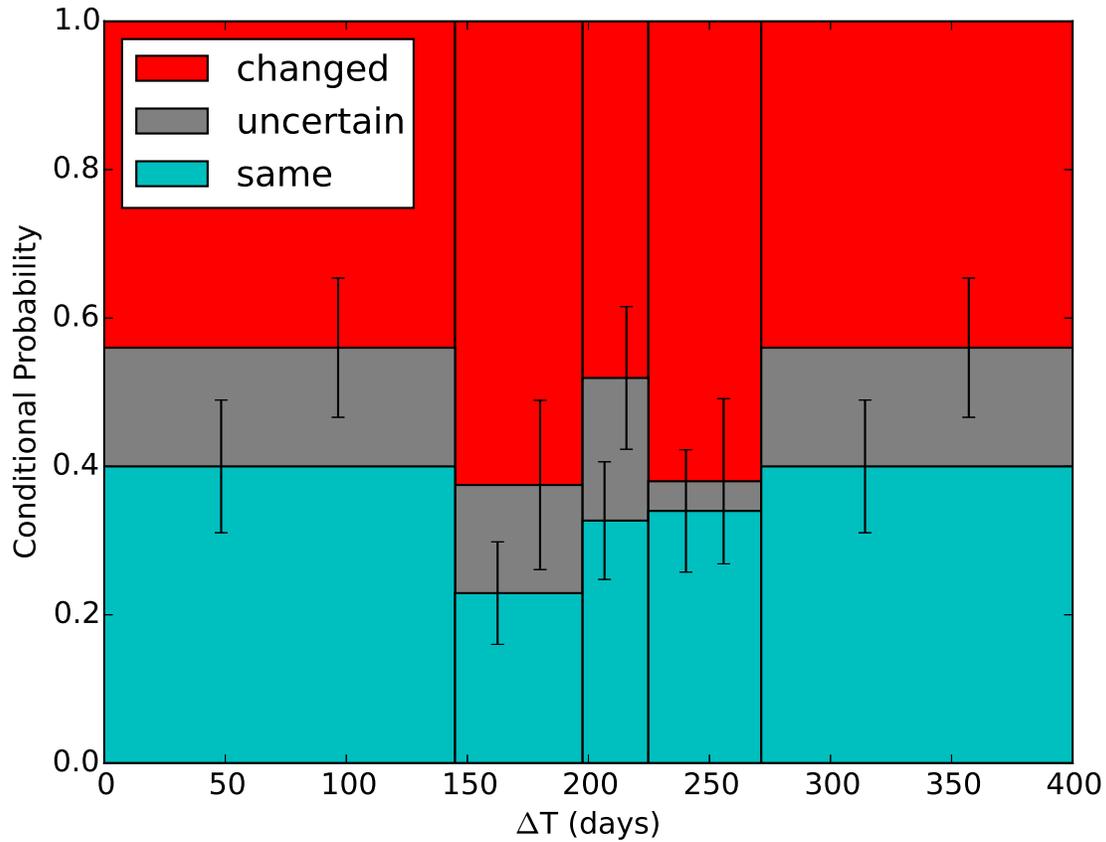}
  \caption{Conditional probability of absorption trend between the 2nd and 3rd epochs continuing in the same direction as observed between the 1st and 2nd epochs, as a function of $\Delta T$ between the 2nd and 3rd epochs.  See text of \S~\ref{sec:conditional} for details.
}
  \label{fig:samflip}
\end{figure}

The result of the above analysis is plotted in Figure~\ref{fig:samflip}.
The plot incorporates three histograms, each with five bins in $\Delta T$,
with the total of all three histograms summing to unity in each bin.
The $\Delta$T on the x-axis is the time frame between the second and third
epochs in the analysis (see Step 4). The histogram widths were
chosen to place 50 measurements in each bin (in an effort standardize the number
statistics for each bin).
The cyan histogram shows the number of troughs that
continued to increase in EW after an increase was already observed ('same').
The red histogram shows the number of troughs that switched their direction of variability ('changed'). 
The grey histogram shows the cases where the direction of variability was unclear ('uncertain').

Within the uncertainties, we are unable to tell the difference between the red
and cyan histograms.
On the time-scales between our 2nd and 3rd epochs,
quasars are equally likely to continue increasing/decreasing
or to stay the same.
Thus, the time-scale between two observations such that the first observation
can predict the second observation (the coherence time-scale) must be less
than 150 days, based on bin-size in Figure~\ref{fig:samflip}.
This result is consistent with the results found by \cite{2015ApJ...806..111G},
using spectroscopic monitoring of one BAL quasar on timescales $<$150 days,
and by \cite{2010ApJ...713..220G},
who found that BAL variability on month-long timescales was not predictive
of BAL variability on multi-year timescales.
Our results are also consistent with the random-walk model for the evolution of
BAL troughs presented in \S 5.4 of \cite{2013ApJ...777..168F}, in which BAL
trough EW variability is modeled as occurring in uncorrelated steps
every $140 \pm 50$ days, on average.

\begin{figure}[htb]
  \centering
  \includegraphics*[scale=0.43]{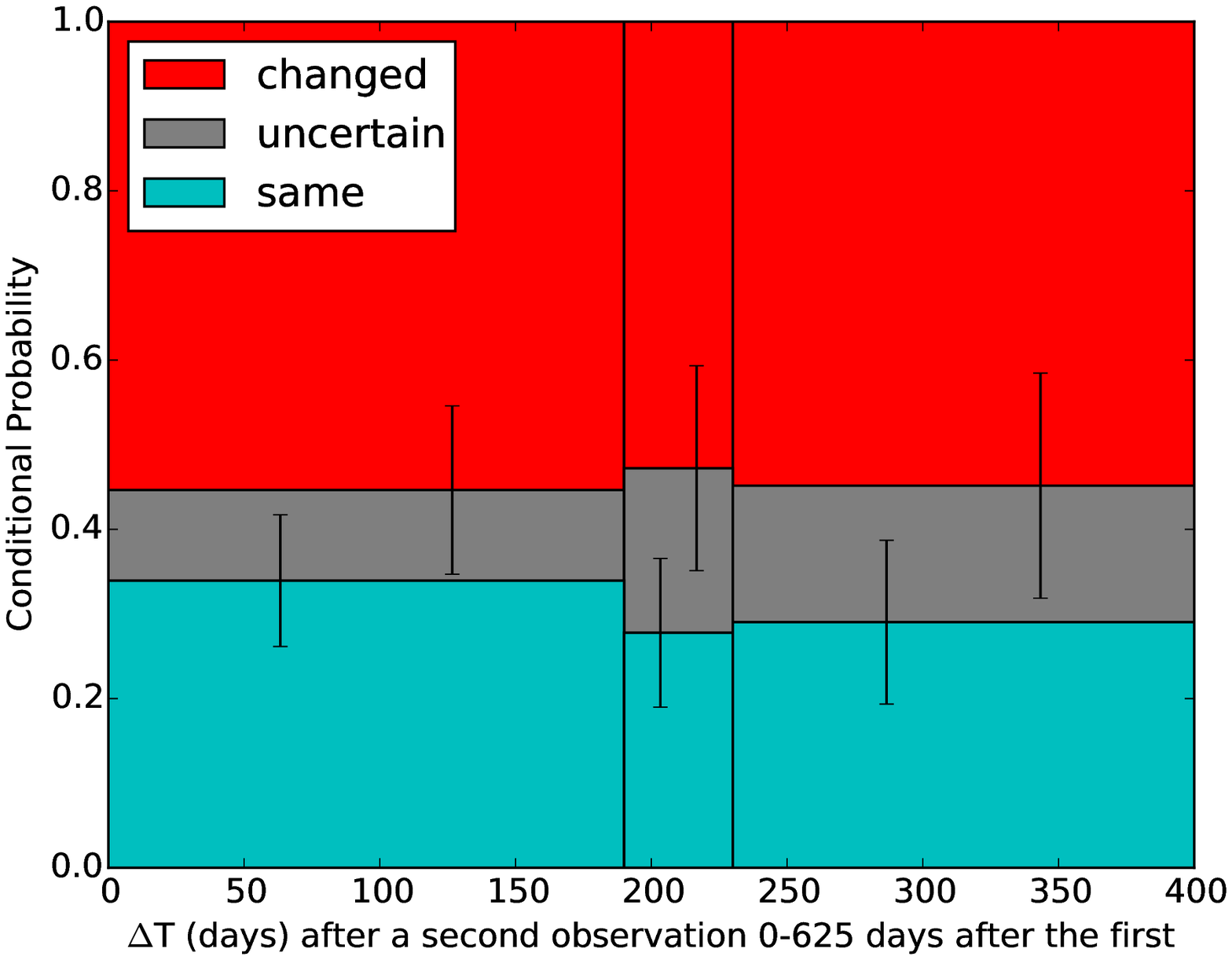}
  \includegraphics*[scale=0.43]{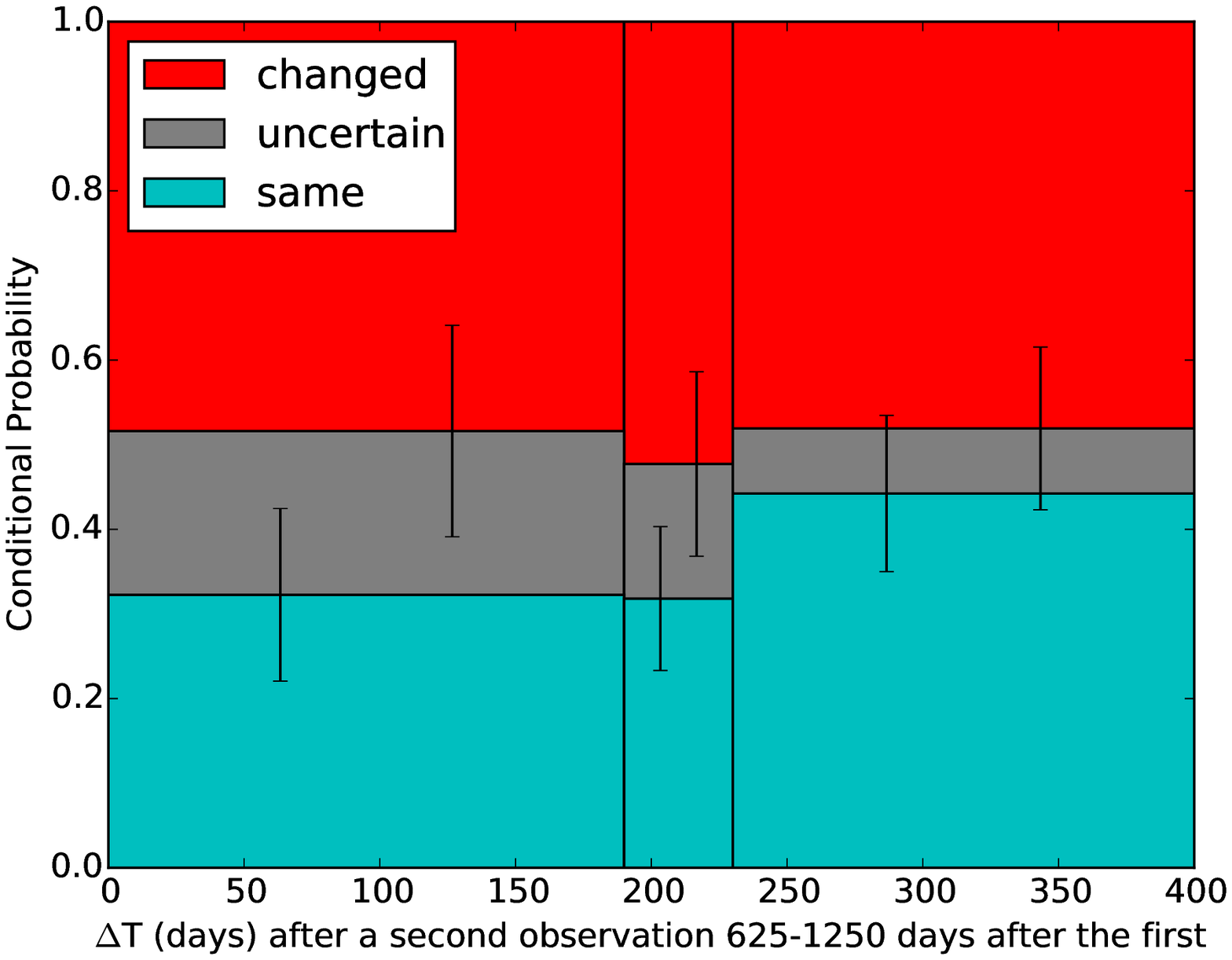}
  \caption{Same as Figure \ref{fig:samflip}, except split into one panel for observations where the $\Delta T$ between the 1st and 2nd epochs was less than 650 days and one panel where it was greater than 650 days.  There is no significant difference between the two cases.}
\label{fig:samflipAB}
\end{figure}

These results remain unchanged if we distinguish between
observations with smaller and larger values of $\Delta T$ between the first two spectroscopic epochs.  To make that comparison, we
make separate conditional probability histograms for observations with
$\Delta T < 625$ days and with $625 < \Delta T < 1250$ days,
using only three bins in
$\Delta T$ between the second and third spectroscopic epochs for each histogram.
The results are shown in Figure \ref{fig:samflipAB}.  No significant difference is seen between those histograms and that shown in Figure \ref{fig:samflip}.

In each absorption complex in each quasar, the EW trend over three epochs can vary in one of
three ways: it can stay the same direction (whether increasing or decreasing),
change direction, or be uncertain.
We have examined the frequency with which all complexes in the same quasar
behave the same way in our observations.
In 15 objects (5 with more than one complex) the EW trend in all complexes present stays the same,
in 31 objects (15 with more than one complex) the EW trend in all complexes present changes direction,
and in 8 objects (2 with more than one complex) all complexes present show uncertain changes.
The remaining objects have complexes that do not all change in the same way; about two-thirds of all objects with more than one complex fall into this category.
Targets observed with sufficiently short time differences between epochs
(and at sufficiently high SNR)
are in principle be more likely to show EW trends that stay the same.
The time differences between epochs in our study are not short enough
for that effect to be seen:
K-S tests reveal no statistical difference between any of the above subsamples'
$\Delta t_{12}$, $\Delta t_{23}$, or $\Delta t_{13}$ distributions.

\subsection{Coordinated Variability}\label{sec:coordvar}

Many of the quasars in our dataset have more than one trough of the same species
varying in their spectra. Comparing how the troughs vary with respect to one another
can help distinguish models of outflows. For example, if two troughs from the same
species in the same
quasar both increase their EW at the same time, or decrease their EW at the
same time, it indicates the source of the variability is affecting different
outflowing absorbers similarly. This is called coordinated variability. Two troughs
could also vary opposite to each other: when one increases in EW, the other
decreases. This is called anticoordinated variability.

It would be difficult to explain coordinated variability in the context
of the transverse motion of clouds across the line of sight to the quasar.
A more natural explanation is the total ionizing flux incident upon the
various individual troughs (from the central source) has increased or decreased
and is thereby responsible for global changes in the absorption, regardless of
outflow velocity of the trough. Thus the most likely cause of coordinated variability in BAL troughs would be a result of changes in the ionizing flux incident upon the absorber.
Coordinated variability has been observed and interpreted previously in e.g.,
\citet{2012ApJ...757..114F}, \citet{2013ApJ...777..168F}, and \citet{wywf}.

To determine whether coordinated
variability was occurring in our dataset, we took all quasars with more
than one absorption complex and calculated the change in EW between
successive epochs for each complex.
A complex was considered to be increasing/decreasing in
absorption if the change in EW was larger than 3 times the statistical noise
propagated from the EW uncertainties.
We then compared each complex's direction of change of EW to that of each other absorption complex in the quasar.

Table~\ref{tab:coordvar} gives the conditional probabilities of how a complex changed given the condition set by another complex in the quasar.
From the
table, if an absorption complex EW was measured to be increasing, then each
other absorption complex EW in that quasar has a 69.8~\%\ chance of increasing
over the same time frame (300 out of the 430 such cases).
If an absorption complex
EW was measured to be decreasing, each other absorption complex EW in that
quasar has a 65.9~\%\ chance to also be decreasing (182 out of 276
such cases).
Both of these situations are considered coordinated variability.
Overall, if a statistically significant change is seen in one complex in a quasar,
there is a $(300+182)/(276+430)=68.3$~\%\ rate of coordinated variability
in other complexes in the same quasar.
The rate of anticoordinated variability is $(70+70)/(430+276)=19.8$~\%.
Only in 11.9~\%\ of cases does a complex not vary statistically significantly
when another complex in the quasar does.
%
We outlined in \S~4.7 of \citet{2013ApJ...777..168F}
how to derive the minimum fraction of observed variability arising from a mechanism that produces coordinated variability.
Random chance will produce an equal mixture of coordinated and
anticoordinated variability, so that
some apparently coordinated variations will be due to chance. To correct for
that effect, we can take the rate of anticoordinated variability and subtract
it from the rate of coordinated variability to estimate the fraction of
BAL EW variability arising from a coordinated process.
The result is that a minimum of 48.5~\%~$\pm$~3.5~\%\ of BAL variability must
be due to a mechanism that produces coordinated variability
(see also He et al. 2017).

\begin{figure}[htb]
  \centering
  \includegraphics[width=0.75\columnwidth]{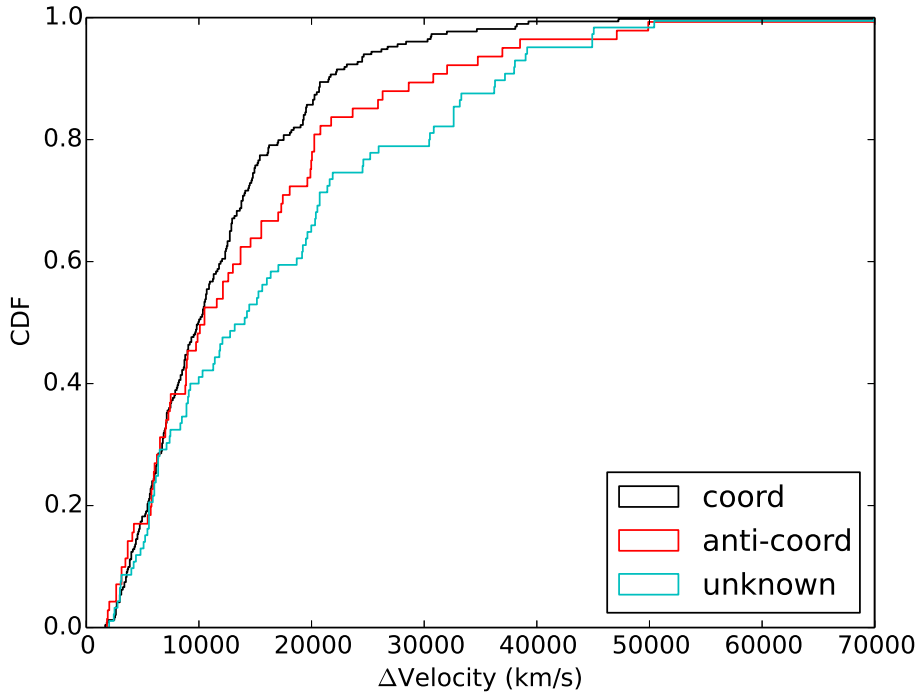}
  \caption{The cumulative distribution functions (CDFs) of coordinated
  variability (black) versus anticoordinated variability (red). For these CDFs, each absorption complex from a given quasar was compared to every
  other complex, which results in double counting.}
  \label{fig:cdfcoordtop}
\end{figure}

With such a strong signal of coordinated variability, we
investigated whether velocity separation between absorption complexes had an
impact on the conditional probabilities of the other complexes.
In Figure~\ref{fig:cdfcoordtop},
the cumulative distribution functions (CDFs) of
coordinated variability (black) versus anticoordinated variability (red) are
plotted as a function of difference in centroid velocity. Also plotted is the
distribution of cases where no statistically significant change (within
the uncertainties) occurred from one observation to the next
in a given complex; this is labeled unknown and plotted in cyan.
Applying the two-sample Kuiper test
to the coordinated/anticoordinated distributions results in a probability of
1.7~\%\ that the two distributions are the same.
Thus, the distributions of velocities for coordinated and
anticoordinated variability are statistically different at 98.3~\% probability. This again suggests there must be
two different mechanisms producing coordinated and
anticoordinated variability.

The incidence of anticoordinated variability increases at larger velocity separations.
In other words, the farther apart troughs are in velocity space,
the less likely they are to have coordinated variability.

Applying the Kuiper test to compare the coordinated and unknown CDFs
results in a probability of $\sim10^{-4}$ that the two histograms are drawn from the same distribution. The coordinated histogram is
different from the unknown histogram at high significance. Finally, the Kuiper test
comparing the anticoordination versus the CDF results in a probability the
two distributions are the same of 15~\%. We cannot
reject the hypothesis that the anticoordination and the unknown CDFs
are drawn from the same sample.

\begin{table}
  \centering
  \caption{The conditional probability, given a change in EW for one absorption complex in a quasar, for what other complexes in the same quasar were doing.
  The first column indicates the condition set by some complex, and the following
  columns indicate the probabilities of other complexes in a quasar varying
  in specific ways. `Increase' is a complex which increased in EW between two
  observations, `same' is a complex which has not changed within the statistical
  uncertainties, and `decrease' is a complex which has decreased in EW between
  two observations.}
  \label{tab:coordvar}
  \begin{tabular}{l|lll|l}
    \hline
    condition & increase & same & decrease & total\\
    \hline
    increase & 69.8\%\ (300) & 13.9\%\ (60) & 16.3\%\ (70) & 430\\
    same & 60\%\ (60) & 16\%\ (16) & 24\%\ (24) & 100 \\
    decrease & 25.4\%\ (70) & 8.69\%\ (24) & 65.9\%\ (182) & 276 \\
    \hline
  \end{tabular}
\end{table}

It is clear there is coordinated variability happening in the dataset.
To investigate whether the relative locations in velocity space of coordinated/anticoordinated complexes matter, we performed a more detailed analysis.
Absorption complexes
were only compared to other complexes at higher velocities,
and we retained the information on velocity separation, $\Delta$T between
observations, and the direction (increasing or decreasing) of the variability.
We followed these steps for each quasar:
\begin{enumerate}
\item{For a given quasar, collect all absorption information for each complex for all
spectra available,
except that we removed comparisons between SDSS1 and SDSS2 observations due to the higher noise levels in both epochs making $\Delta EW$ measurements very uncertain.}
\item{Sort the complexes in ascending velocity order (i.e., closest to \civ\
emission to furthest).}
\item{Compare first complex to second, third, ..., $n$th. For each comparison record direction of coordination (see below), separation in centroid velocity, and $\Delta$T between observations.}
\item{Compare second complex to third, ..., $n$th. For each comparison record direction of coordination (see below), separation in centroid velocity, and $\Delta$T between observations.}
\item{Repeat until no more absorption complexes to compare in the quasar.}
\end{enumerate}
The results of this analysis are given in Table~\ref{tab:coordvarbetter}.
The numbers of cases considered is one-half of those considered in the
previous table because in the above analysis we compared complexes to others
at both lower and higher velocities.

As was argued above, if we assume that all comparisons between two
troughs will give a mixture of both coordinated and anticoordinated
variability, and thus remove the effect of anticoordinated variability (19.3~\%) from the rate of coordinated variability (66.6~\%), there is
at minimum a 47.3~\%\ rate of coordinated
variability motivated by some underlying mechanism and not by chance.

\begin{figure}[htb]
  \centering
  \includegraphics[width=0.75\columnwidth]{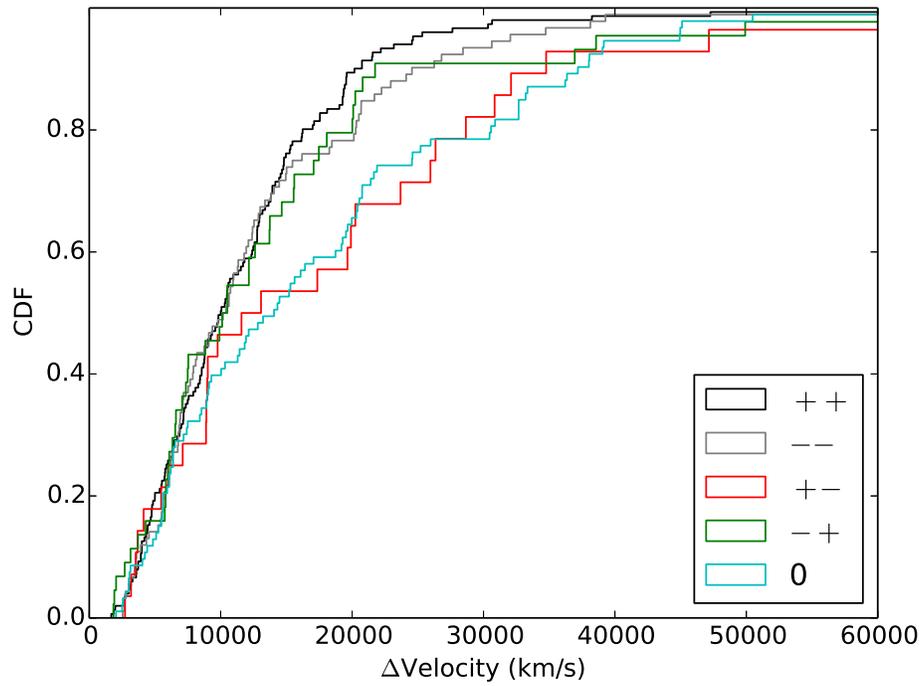}
  \caption{This plot
  removes double counting and separates out the different cases of
  coordination/anticoordination. Cases are indicated by:
  `$++$' when two troughs increased in EW together,
  `$--$' when two troughs decreased in EW together,
  `$-+$' when the lowest-velocity trough decrease in EW while the other
  increased,
  `$+-$' when the lowest-velocity trough increased in EW while the other
  decreased,
  and finally `0' for all cases that involved a change that was below
  the 3$\sigma$ statistical noise limit.}
  \label{fig:cdfcoordbottom}
\end{figure}

In Figure~\ref{fig:cdfcoordbottom} are the cumulative
distribution functions for the data in Table~\ref{tab:coordvarbetter}.
In the Figure, we have
separated out the differing possibilities for how troughs can respond with
respect to one another. The symbol `$++$' indicates the case when two troughs
have increased in absorption at the same time, while `$--$' indicates both
decreased. The two anticoordination cases are `$-+$' and `$+-$'; the former is
when the lower-velocity absorption complex is decreasing in EW
while the higher-velocity absorption complex is increasing,
and the latter is when the higher-velocity absorption complex is decreasing
while the lower-velocity absorption complex is increasing in EW.
All cases in which no statistically significant change was measured in either or both troughs are labeled `$0$'.

It is clear that both cases of coordinated variability are similar to each
other; the Kuiper test yields
a 42~\%\ probability that both CDFs are drawn from the same sample, meaning
there is no favoured direction of coordinated variability.
While in the figure there is an obvious difference between the
`$-+$' and `$+-$' cases of anticoordinated variability, the Kuiper test between these two cases indicates a probability of 56~\% that those two CDFs are actually drawn from the same sample.
Small-number statistics mean that the apparent difference between the `$+-$'
and `$-+$' histograms, while intriguing, cannot be interpreted as real until
better statistics are obtained.


Both Figure~\ref{fig:cdfcoordtop} and \ref{fig:cdfcoordbottom} show that coordinated variability is happening, and the troughs closer together in velocity are more likely to vary in a coordinated way.
We speculate that this result arises from density considerations.
Higher velocity outflows are likely to have lower densities due to mass conservation in an accelerating flow.
Since the ionization response time of an absorbing cloud is related to its density, it is plausible that two outflows in the same quasar with a large separation in velocity would be less likely to vary in unison.
The density in an accelerating flow with velocity $v(r)$ satisfies the continuity equation $n(r)A(r)v(r)={\rm constant}$, where $A(r)$ is the cross-sectional area of the flow as a function of distance $r$ from the launch point.  If $A(r)$ is constant, then $n \propto 1/v$.  For example, Figure 4 of \cite{mcgv} shows that $n$ drops by a factor of $10^3$ as the velocity increases by a factor of $10^3$ beyond the sonic point of a disk wind outflow, and then $n$ drops further as the gas coasts to larger radii at its terminal velocity (constant $v$ but increasing $A$).
The timescale needed for gas to respond to ionizing flux changes can often be approximated as $t\propto 1/n\alpha$ where $\alpha$ is the relevant recombination coefficient \cite{2015ApJ...806..111G}.
Thus, in clumps expanding in an accelerating flow, gas at two different outflow velocities will have a ratio of response timescales $t_{hi}/t_{lo} > v_{hi}/v_{lo}$ for $v_{lo}>10$ km s$^{-1}$ (an estimate of the turbulent sound speed).
Our sample of absorbers at 1000 to 60,000 km s$^{-1}$ could span a factor of $>60$ range in response timescales, and even a factor of 10 is sufficient to show anticoordination in response to the same underlying ionizing flux (Figure 4 of \citealt{2012A&A...544A..33A}).

\begin{table}
  \centering
  \caption{The conditional probability, given a change in EW for one absorption complex in a
  quasar, for what other complexes at higher velocities in the same quasar were doing.  The format is the same as for Table \ref{tab:coordvar}.}
  \label{tab:coordvarbetter}
  \begin{tabular}{l|lll|l}
    \hline
    condition & increase & same & decrease & total\\
    \hline
    increase & 69.4\%\ (150) & 18.1\%\ (39) & 12.5\%\ (27) & 216\\
    same & 51.2\%\ (21) & 19.5\%\ (8) & 29.3\%\ (12) & 41 \\
    decrease & 29.4\%\ (43) & 8.2\%\ (12) & 62.3\%\ (91) & 146 \\
    \hline
  \end{tabular}
\end{table}



\clearpage

\section{Conclusions}\label{sec:conclude}

Below is a summary of the notable conclusions found in this work.

\begin{enumerate}

\item{Visual comparison of multiple spectra of the same quasar from SDSS DR7
and DR9 or DR7 and DR10 yielded
306
visually identified candidates for newly
emerged BAL troughs. Absorption was quantitatively confirmed in 103 out of the
105 of these cases which were followed up with Gemini spectroscopy.
Thus, the visual inspection is robust.
  See \S~\ref{sec:targetSelection}.}

  \item{Using a modified version of the BALnicity Index, 653 individual
  broad absorption troughs were identified in 360 spectra of the 105 targets
  in our dataset. Each trough was at least 1,000~\kms\ wide and found anywhere
  between $0 < v~(\textrm{\kms}) < $ 65,000, where 0~\kms\ is at the
  systemic redshift of the quasar, and positive velocities are toward
  the observer. See \S~\ref{sec:visinspec}.}

  \item{To account for troughs splitting apart or merging into one between
  epochs, we defined an absorption complex to be a region in a quasar spectrum
  that, over the multiple spectra available, has had one or more BAL trough
  in at least one epoch. The 653 individual troughs are reduced to 219
  absorption complexes by this definition. See \S~\ref{sec:compident}.}

  \item{In the sample of 105 quasars, there were 36 instances of a quasar transitioning
  from non-BAL to BAL. By design, this mostly occurred in the SDSS-BOSS
  transition. There were 11 cases of quasars transitioning from BAL to non-BAL.
  See \S~\ref{sec:baltrans}.}

\item{The visually identified BAL trough emergence rate at $z>1.68$ in our
sample was $1.87 \pm 0.11$~\%\ over rest-frame timescales of $1-3$ years.
The rate of emergence of BAL quasars from objects previously classified as
non-BAL quasars was $0.59 \pm 0.12$~\%.
We find that our BAL quasar emergence rates and the BAL quasar disappearance
rates of \citep{2012ApJ...757..114F} are in agreement for a BAL quasar
fraction in the parent sample of $f_{BAL}=0.26^{+0.09}_{-0.12}$.
See \S~\ref{sec:emergenceRate}.}

\item{We find that candidate newly emerged BAL quasar troughs are preferentially drawn from among BAL troughs with smaller balnicity indices, shallower depths, larger velocities, and smaller widths. See \S~\ref{sec:DR12QBAL}.}

  \item{As expected, the weighted-mean change in absorption equivalent width $\Delta$EW from SDSS to BOSS was
  $4.16\pm0.10$~\AA, an increase in EW. After the SDSS-BOSS transition, however,
  there was a significant trend for the EW to decrease. The weighted-mean $\Delta$EW from
  BOSS-GEM was $-0.93\pm0.09$~\AA. See \S~\ref{sec:afteremerge}.}

  \item{Despite the above averages, a more detailed analysis that took into
  account the time between observations found: if an absorption complex begins
  to increase in EW between two observations, it is equally likely to continue
  increasing in the next observation, as it is to switch to decreasing.
  Based on our analysis this indicates the coherence time-scale of BAL EW
  variations must be less than 150 days. See \S~\ref{sec:afteremerge}.}

  \item{Coordinated variability, when two absorption complexes in the same quasar increase or decrease in EW at the same time, occurs at a rate of 68.3~\%. Anticoordinated variability occurs at a rate of 19.8~\%. This likely indicates
  that ionizing flux variations contribute to at least
  48.5~\%~$\pm$~3.5~\%\ of all BAL EW variability. See \S~\ref{sec:coordvar}.}

  \item{Troughs that are closer together in velocity space are more likely to exhibit coordinated variability than those further apart in velocity space, at
98.3~\% significance.
This dependence of coordinated variability
  on velocity separation may be related to cloud densities.
  If clouds decrease in density as they accelerate, higher-velocity clouds would respond to changes in the ionizing flux differently than lower-velocity clouds.
There is also a suggestion of different velocity distributions for cases of a lower velocity trough increasing in EW while a higher velocity trough decreases in EW as compared to the reverse; however, small number statistics prevent a determination of whether the suggested difference is statistically significant or not.
See \S~\ref{sec:coordvar}.}

  \item{The change in equivalent width, $\Delta$EW, with respect to time
  between successive observations in our data merges smoothly with the distribution seen in the literature.
  Our data filled in the region on rest-frame time-scales of $0.5-1.0$ years that been largely unexplored in the literature.
  See Figure~\ref{fig:deltaEW}.}

\end{enumerate}

\subsection{Future Directions}

In \S~\ref{sec:afteremerge}, we attempted to determine if a BAL's past
variability was a predictor of its variability in the future.
Unfortunately, our dataset did not have the number statistics, nor the time
resolution to conclude that it could. Certainly one can imagine that if a
trough were to be observed from zero EW to a full trough with multiple spectral
epochs, it would be possible to use a history of increasing strength of absorption
to predict
the trough
will continue to increase with some constraint on time.
In our dataset, that coherence
time-scale must be less than 100 days in the rest-frame.
Performing a monitoring program with high time resolution for a medium-sized
set of quasars would help determine a coherence scale. For example, a set of 10
bright BAL quasars observed once a week for an entire observing semester would
provide enough data to answer this question. It would be recommended to choose
a set of quasars from the dataset presented herein, due to the already
confirmed variable absorption in them.

A monitoring program focused on high time resolution could inform understanding
on the nature of coordinated variability as well. We observed an apparent
difference between two types of anticoordinated variability that would
challenge our understanding of the quasar model if it were proven correct.
The only way to determine its validity would be a campaign of observations on
quasars with multiple troughs with higher time resolution.

In the dataset presented here, a few quasars presented
very complicated and informative variability. Our separately published case study of J023011 \citep{J0230} is an example of how targeted studies can lead to more specific results.
In that study, we showed how the observed short-timescale absorption variability in J023011 could be used to constrain one of the absorber's physical parameters, depending on the cause of EW variability: its electron density in the case of ionization balance changes, or its transverse velocity in the case of bulk motion of gas across the line of sight.
To that end, continued monitoring of quasar J015017
(see Figure~\ref{fig:j015017}) could help disentangle the different signals of ionization changes and bulk motion.



\acknowledgments

The authors wish to recognize and acknowledge the very significant cultural role and reverence that the summit of Maunakea has always had within the indigenous Hawaiian community.  We are most fortunate to have the opportunity to conduct observations from this mountain.
We thank the referee for their comments.
P.B.H., J.A.R., and P.R.H. were supported in part by the Natural Sciences and Engineering Research Council of Canada (NSERC), funding reference number 2017-05983, and by an Ontario Early Researcher Award to P.B.H.
N.F.A. would like to acknowledge financial support from TUBITAK (115F037).



\vspace{5mm}
\facilities{Sloan, Gemini:South, Gemini:North}

\software{IRAF, IDL, python}

\listofchanges

\end{document}